\documentclass[a4paper,12pt]{article}
\pdfoutput=1
\usepackage{amssymb}
\usepackage{amsmath}
\usepackage{amsfonts}
\usepackage{mathtools}
\usepackage{bm}
\usepackage{hyperref,graphicx}
\usepackage{slashed}
\usepackage{color}
\usepackage{indentfirst}
\usepackage{graphicx}
\usepackage{bbm}
\usepackage{csquotes} 
\usepackage{caption}
\usepackage{here}
\usepackage{comment}
\usepackage{listings}
\usepackage[italicdiff]{physics}
\numberwithin{equation}{section} 

\usepackage[
backend=biber,
eprint=true,
sorting=none,
sortcites=true,
maxnames=8,
date=year
]{biblatex}
\def\ignore#1{{}}

\makeatletter
\newcommand*\rel@kern[1]{\kern#1\dimexpr\macc@kerna}
\newcommand*\widebar[1]{%
  \begingroup
  \def\mathaccent##1##2{%
    \rel@kern{0.8}%
    \overline{\rel@kern{-0.8}\macc@nucleus\rel@kern{0.2}}%
    \rel@kern{-0.2}%
  }%
  \macc@depth\@ne
  \let\math@bgroup\@empty \let\math@egroup\macc@set@skewchar
  \mathsurround\z@ \frozen@everymath{\mathgroup\macc@group\relax}%
  \macc@set@skewchar\relax
  \let\mathaccentV\macc@nested@a
  \macc@nested@a\relax111{#1}%
  \endgroup
}
\makeatother

\renewcommand{\thefootnote}{\arabic{footnote}}
\clearpage
\addtocounter{page}{-1}

\def\ignore#1{{}}

\newcommand{\alp}{\alpha}
\newcommand{\bt}{\beta}

\newcommand{\Gm}{\Gamma}
\newcommand{\dlt}{\delta}
\newcommand{\Dlt}{\Delta}

\newcommand{\tht}{\theta}

\newcommand{\lmd}{\lambda}

\newcommand{\sgm}{\sigma}

\newcommand{\vph}{\varphi}

\newcommand{\bea}{\begin{eqnarray}}
\newcommand{\eea}{\end{eqnarray}}

\newcommand{\tl}[1]{\tilde{#1}}

\newcommand{\der}{\partial}

\newcommand{\id}{\mbox{\boldmath $1$}}

\newcommand{\brkt}[1]{\left( #1 \right)}
\newcommand{\brc}[1]{\left\{ #1 \right\}}
\newcommand{\sbk}[1]{\left[ #1 \right]}

\renewcommand{\Re}{{\rm Re}\,}
\renewcommand{\Im}{{\rm Im}\,}

\newcommand{\cE}{{\cal E}}
\newcommand{\cF}{{\cal F}}
\newcommand{\cG}{{\cal G}}
\newcommand{\cH}{{\cal H}}

\newcommand{\cN}{{\cal N}}

\newcommand{\cR}{{\cal R}}
\newcommand{\cS}{{\cal S}}

\newcommand{\cU}{{\cal U}}
\newcommand{\cV}{{\cal V}}

\newcommand{\Tmp}{T_{\rm tmp}}
\newcommand{\tref}{t_{\rm ref}}

\newcommand{\bAref}{\bar{A}_{\rm ref}}
\newcommand{\btref}{\hat{\beta}_{\rm ref}}
\newcommand{\btmod}{\hat{\beta}_{\rm mod}}
\newcommand{\Cmod}{C_{\rm mod}}
\newcommand{\Crad}{C_{\rm rad}}
\newcommand{\Rmr}{R^{\rm mod}_{\rm rad}}
\newcommand{\trad}{t_{\rm rad}}
\newcommand{\tmod}{t_{\rm mod}}
\newcommand{\tlf}{t_{\rm dc}}
\newcommand{\Gmm}{\Gamma_{\rm mod}}
\newcommand{\tc}{t_{\rm c}}
\newcommand{\tlCmod}{\tilde{C}_{\rm mod}}
\newcommand{\Drad}{D_{\rm rad}}
\newcommand{\CA}{C_A}

\renewcommand{\thefootnote}{\fnsymbol{footnote}}

\begin{document}

\title{
\begin{flushright}
\begin{minipage}{0.25\linewidth}
\normalsize
KEK-TH-2603 \\
KYUSHU-HET-282 \\*[50pt]
\end{minipage}
\end{flushright}
{\Large \bf 
Induced moduli oscillation by radiation and space expansion in a higher-dimensional model
\\*[20pt]}}

\author{
Hajime~Otsuka$^{a}$\footnote{
E-mail address: otsuka.hajime@phys.kyushu-u.ac.jp
}
\ and\
Yutaka~Sakamura$^{b,c}$\footnote{
E-mail address: sakamura@post.kek.jp
}\\*[20pt]
$^a${\it \normalsize
Department of Physics, Kyushu University,}\\ 
{\it \normalsize
744 Motooka, Nishi-ku, Fukuoka, 
819-0395, Japan}\\
$^b${\it \normalsize 
KEK Theory Center, Institute of Particle and Nuclear Studies, KEK,}\\
{\it \normalsize 1-1 Oho, Tsukuba, Ibaraki 305-0801, Japan}\\
$^c${\it \normalsize 
Graduate University for Advanced Studies (Sokendai),}\\
{\it \normalsize 1-1 Oho, Tsukuba, Ibaraki 305-0801, Japan.}
}
\date{}

\maketitle

\centerline{\small \bf Abstract}
\begin{minipage}{0.9\linewidth}
\medskip 
\medskip 
\small

We investigate the cosmological expansion of the 3D space in a 6D model compactified on a sphere, 
beyond the 4D effective theory analysis. 
We focus on a case that the initial temperature is higher than the compactification scale. 
In such a case, the pressure for the compact space affects the moduli dynamics 
and induces the moduli oscillation even if they are stabilized at the initial time. 
Under some plausible assumptions, we derive the explicit expressions for the 3D scale factor 
and the moduli background in terms of analytic functions. 
Using them, we evaluate the transition times between different cosmological eras 
as functions of the model parameters and the initial temperature. 
\end{minipage}

\renewcommand{\thefootnote}{\arabic{footnote}}
\thispagestyle{empty}
\clearpage

\section{Introduction}
\label{introduction}

The existence of the extra dimensions is predicted in string theory. 
Since the experimental constraints on the size of the extra dimensions that the gravity feels 
are much weaker than those for the standard model particles, there is a wide allowed parameter space 
for brane-world models with relatively large extra dimensions~\cite{Arkani-Hamed:1998jmv,Arkani-Hamed:1998sfv,Antoniadis:1998ig}. 
This type of models have been considered not only as a solution to the large hierarchy problem, 
but also as a solution to the cosmological constant problem~\cite{Aghababaie:2003wz}. 
Recently, it is also pointed out that 
a micron-size large extra dimension may be predicted by the swampland conjecture~\cite{Lust:2019zwm},  
which is called the dark dimension scenario~\cite{Montero:2022prj}. 
If a large extra compact space exists, it affects the cosmological history at early times. 
In particular, the temperature~$\Tmp$ in the radiation-dominated era can be higher than the compactification scale~$m_{\rm KK}$  
since the latter has a small value in such a case. 

In our previous papers~\cite{Otsuka:2022rpx,Otsuka:2022vgf}, 
we studied the time evolution of the space and the background values of the moduli 
in a six-dimensional (6D) model compactified on a sphere~$S^2$ by solving the 6D field equations numerically.\footnote{
This model is basically based on a gauged 6D supergravity on $S^2$~\cite{Nishino:1984gk}, 
which has many interesting properties, e.g., 
a self-tuning of the four-dimensional (4D) vacuum energy~\cite{Aghababaie:2003wz,Burgess:2004kd,Burgess:2004dh,Garriga:2004tq} 
and a verification of swampland conjecture~\cite{Anchordoqui:2020sqo}. 
Its string realization is also discussed in Ref.~\cite{Cvetic:2003xr}.  
} 
We found that when the initial temperature of the universe is higher than $m_{\rm KK}$, 
the expansion rate for the three-dimensional (3D) non-compact space deviates 
from that of the usual 4D cosmology. 
In general, the 3D scale factor~$e^A$ evolves as $t^{2/(3(1+w))}$, where $t$ is the cosmological time and 
$w$ is the ratio of the pressure to the energy density. 
In the radiation-dominated era, $w^{-1}$ measures the dimensions that the radiation feels. 
In fact, when $\Tmp>m_{\rm KK}$, the radiation feels the whole five-dimensional (5D) space 
and $e^A\propto t^{5/9}$. 
As the universe expands and the temperature goes down, the radiation gradually ceases to feel 
the compact space, and $w^{-1}$ approaches to three after $\Tmp$ gets lower than $m_{\rm KK}$. 
Then, the expansion rate slows down to $e^A\propto t^{1/2}$. 
We also found that even if the moduli are stabilized at the initial time, the moduli oscillation is induced 
by the pressure for the two-dimensional (2D) compact space~$p_2^{\rm rad}$. 
This effect cannot be discussed in the conventional 4D effective theory approach 
since $p_2^{\rm rad}$ is absent in the 4D Einstein equation. 
When $\Tmp>m_{\rm KK}$, the effect of $p_2^{\rm rad}$ on the moduli dynamics 
cannot be neglected. 
If the lifetime of the moduli is long enough, 
the induced moduli oscillation eventually dominates the energy density and the 3D space expands as $e^A\propto t^{2/3}$ 
at later times. 
Therefore, there are the following eras in this setup. 
\begin{enumerate}
 \item 6D radiation-dominated era ($e^A\propto t^{5/9}$)
 \label{6rad}

 \item 4D radiation-dominated era ($e^A\propto t^{1/2}$)
 \label{4rad}
 
 \item (Induced) moduli-oscillation-dominated era ($e^A\propto t^{2/3}$)
 \label{mod_osc_era}
\end{enumerate}
The era~\ref{mod_osc_era} will end by the decay of the moduli, and transition into 
the 4D radiation-dominated era again~\cite{Moroi:2001ct}. 
After that, the universe behaves as the standard cosmology. 
Let us denote the transition time from the era~\ref{6rad} to the era~\ref{4rad} as $\trad$, 
and that from \ref{4rad} to \ref{mod_osc_era} as $\tmod$. 
In principle, the spacetime evolution is determined once the model parameters and the initial conditions are provided. 
However, since these results are obtained by the numerical computations in the previous works, 
we cannot directly see how the transition times~$\trad$ and $\tmod$ depend on the initial parameters. 
Besides, it is difficult to pursue the whole history of the universe 
due to the limitation of the computational power. 

In this paper, we derive approximate expressions for the 3D scale factor, the moduli background values and the transition times 
in terms of analytic functions by solving the 6D evolution equations under some approximations. 
Since we can discriminate the eras by the power~$p$, which is defined as $e^A\propto t^p$ in each era, 
we will focus on the change of $p$ during the spacetime evolution.  
The expressions derived in this paper enable us to pursue the spacetime evolution until much later times than the previous works, 
and to clarify the dependence of the transition times~$\trad$ and $\tmod$ 
on the model parameters and the initial temperature. 
These results will help make discussions transparent. 

The paper is organized as follows. 
In Sec.~\ref{setup}, we briefly explain our setup and show the evolution equations. 
In Sec.~\ref{space_evolve}, we derive analytic expressions for various quantities 
by solving the evolution equations under some plausible approximations. 
We then define the effective power~$p$ and derive its explicit expression 
using the functions we have defined. 
In Sec.~\ref{evolve:p}, we discuss the time evolution of $p$, and estimate the transition times. 
Sec.~\ref{conclusion} devoted to the summary. 
In Appendix~\ref{TDquantities}, brief derivations of the energy density and the pressures for the radiation 
are shown. 
In Appendix~\ref{conserv_law}, we derive the evolution equation for the temperature 
from the conservation law of the energy-momentum tensor. 
In Appendix~\ref{forced_osc}, we provide a general solution to the inhomogeneous differential equation  
that describes the moduli oscillation.

\section{Setup}
\label{setup}
We consider a 6D model used in our previous works~\cite{Otsuka:2022rpx,Otsuka:2022vgf}. 
In this section, we briefly review the model and the evolution equations for the universe. 
The spacetime is compactified on a 2D sphere~$S^2$. 
As coordinates on $S^2$, we choose the spherical ones~$(x^4,x^5)=(\theta,\phi)$, 
where $\tht$ and $\phi$ are the polar and the azimuthal angles, respectively. 

The action is given by~\footnote{
Throughout the paper except for Secs.~\ref{transition_times} and \ref{moduli_decay}, 
we work in the 6D Planck unit~$M_6=1$, where $M_6$ is the 6D Planck mass. 
} 
\begin{align}
 S &= \int d^6x\;\sqrt{-g^{(6)}}\brc{-\frac{1}{2}R^{(6)}-\frac{1}{2}\der^M\sgm\der_M\sgm-\frac{g_{\rm gc}^2e^\sgm}{4}F^{MN}F_{MN}-V_{\rm pot}(\sgm)}, 
 \label{6Daction}
\end{align}
where $M,N=0,1,\cdots,5$ denote the 6D indices, $g^{(6)}$ is the determinant of the 6D metric tensor, 
$R^{(6)}$ is the 6D Ricci scalar, $\sgm$ is a real scalar, $F_{MN}\equiv \der_MA_N-\der_NA_M$ is the field strength of 
a U(1) gauge field~$A_M$, and $g_{\rm gc}$ is the gauge  coupling constant. 
The scalar potential~$V_{\rm pot}(\sgm)$ is given by
\begin{align}
 V_{\rm pot}(\sgm) &= 2e^{-\sgm}+\frac{m^2}{2}\brkt{\sgm-\sgm_*}^2, 
 \label{def:Vpot}
\end{align}
where $m$ and $\sgm_*$ are positive constants. 

Except for the second term in \eqref{def:Vpot}, 
the action~\eqref{6Daction} can be embedded into a gauged 6D $\cN=(1,0)$ supergravity~\cite{Nishino:1984gk,Randjbar-Daemi:1985tdc}. 
We add the second term in order to stabilize the moduli completely. 

In this paper, we neglect effects of the 3-branes, one of which the standard model particles live~\footnote{
See for more details about codimension-two branes~\cite{Gibbons:2003di,Aghababaie:2003ar,Lee:2005az}. 
}, 
and assume that the background spacetime has homogeneity and isotropy for 3D non-compact space and a spherical symmetry for $S^2$. 
Thus, we take the following ansatz for the background fields. 
\begin{align}
 g_{MN} &= \begin{pmatrix} -1 & & & \\ & e^{2A(t)}\id_3 & & \\ & & e^{2B(t)} & \\ & & & e^{2B(t)}\sin^2\theta \end{pmatrix}, \nonumber\\
 F_{\mu\nu} &= F_{\mu\tht} = F_{\mu\phi} = 0, \nonumber\\
 F_{\tht\phi} &= -F_{\phi\tht} = \frac{\sin\theta}{2g_{\rm gc}}, \;\;\;\;\;
 F_{\tht\tht} = F_{\phi\phi} = 0, \nonumber\\
 \sgm &= \sgm(t), 
 \label{def:background}
\end{align}
where $\mu,\nu=0,1,2,3$ are the 4D indices.

In the absence of the radiation in the bulk, the model has the following static solution.\footnote{
Throughout the paper, we normalize the 3D scale factor as $A(t=0)=0$. 
} 
\begin{align}
 A&= 0, \;\;\;\;\;
 B = B_* \equiv \frac{\sgm_*}{2}-\ln 2, \;\;\;\;\;
 \sgm = \sgm_*, 
 \label{static_sol}
\end{align}
and the Kaluza-Klein (KK) mass scale is given by~\footnote{
The mass eigenvalues for the KK gravitons are expressed as $\sqrt{l(l+1)}m_{\rm KK}$, where $l$ is a non-negative integer. 
} 
\begin{align}
 m_{\rm KK} &\equiv e^{-B_*} = 2e^{-\sgm_*/2}. 
 \label{def:mKK}
\end{align} 

In addition to the above field content, we introduce the radiation in the bulk. 
In 6D $\cN=1$ supergravity, the number of hypermultiplets~$n_H$ and that of vector multiplets~$n_V$ are constrained by 
the anomaly cancellation condition~$n_H-n_V=244$~\cite{Randjbar-Daemi:1985tdc,Green:1984bx,Kumar:2010ru}.\footnote{
The number of tensor multiplets is assumed to be one, otherwise the theory cannot be described by the Lagrangian. 
}
Therefore, at least 245 hypermultiplets exist in the bulk. 
Since each hypermultiplet has four bosonic and four fermionic degrees of freedom, 
we assume that the degrees of freedom for the radiation is $g_{\rm dof}=2000$ in this paper. 
Due to the isometries of the spacetime, the energy-momentum tensor for the radiation has the form of 
\begin{align}
 (T^{\rm rad})_M^{\;\;N} &= \begin{pmatrix} \rho^{\rm rad} & & & \\ & -p_3^{\rm rad} \id_3 & & \\ & & -p_2^{\rm rad} & \\ & & & -p_2^{\rm rad}\sin^2\tht \end{pmatrix}, 
 \label{def:T^rad}
\end{align}
where $\rho^{\rm rad}$, $p_3^{\rm rad}$ and $p_2^{\rm rad}$ are the radiation energy density, the pressures in the non-compact 3D space 
and in the compact 2D space, respectively. 
Their explicit forms are listed in Appendix~\ref{TDquantities}. 

In the presence of the radiation, the static solution~\eqref{static_sol} is no longer a solution of the field equations, 
and the universe continues to expand. 
Substituting the background ansatz~\eqref{def:background} and \eqref{def:T^rad} into the 6D Einstein equations and the dilaton field equation, 
we obtain the evolution equations for the background fields, which are summarized as
\begin{align}
 \ddot{A} &= -\brkt{3\dot{A}+2\dot{B}}\dot{A}+\brkt{e^{-\sgm}-\frac{e^{\sgm-4B}}{16}}
 +\frac{m^2}{4}\brkt{\sgm-\sgm_*}^2
 +p_3^{\rm rad}, \nonumber\\
 \ddot{B} &=-\brkt{3\dot{A}+2\dot{B}}\dot{B}+\brkt{e^{-\frac{\sgm}{2}}-\frac{e^{\frac{\sgm}{2}-2B}}{4}}
 \brkt{e^{-\frac{\sgm}{2}}-\frac{3e^{\frac{\sgm}{2}-2B}}{4}}+\frac{m^2}{4}\brkt{\sgm-\sgm_*}^2+p_2^{\rm rad}, \nonumber\\
 \ddot{\sgm} &= -\brkt{3\dot{A}+2\dot{B}}\dot{\sgm}+2\brkt{e^{-\sgm}-\frac{e^{\sgm-4B}}{16}}
 -m^2\brkt{\sgm-\sgm_*}, 
 \label{evolution_eqs}
\end{align}
with the constraint, 
\begin{align}
 3\dot{A}^2+\dot{B}^2+6\dot{A}\dot{B}-\frac{1}{2}\dot{\sgm}^2 &= 2\brkt{e^{-\frac{\sgm}{2}}-\frac{e^{\frac{\sgm}{2}-2B}}{4}}^2
 +\frac{m^2}{2}\brkt{\sgm-\sgm_*}^2+\rho^{\rm rad} \nonumber\\
 &\equiv \hat{\rho}^{\rm tot}. 
 \label{constraint}
\end{align}
We have used the relation~\eqref{rel:rhop}. 

The energy density and the pressures are expressed as (see Appendix~\ref{TDquantities})~\footnote{
To simplify the discussion, we assume that the chemical potential~$\mu$ is negligible, i.e., $\bt\mu\ll 1$, and the radiation consists of bosons and fermions 
with the same degrees of freedom. 
}
\begin{align}
 \rho^{\rm rad} &= \frac{g_{\rm dof}e^{-2B}}{8\pi^3\bt^4}\brkt{\frac{\pi^4}{16}+3Q_1+Q_2}, \nonumber\\
 p_3^{\rm rad} &= \frac{g_{\rm dof}e^{-2B}}{8\pi^3\bt^4}\brkt{\frac{\pi^4}{48}+Q_1}, \nonumber\\
 p_2^{\rm rad} &= \frac{g_{\rm dof}e^{-2B}}{16\pi^3\bt^4}Q_2, 
 \label{expr:rho-p:text}
\end{align}
where $\beta\equiv 1/\Tmp$ is the inverse temperature, 
the functions~$Q_1(x)$ and $Q_2(x)$ are defined in \eqref{def:Q_1} and \eqref{def:Q_2} respectively, 
and their arguments are $e^{-B}\bt$. 
The evolution equation for $\bt$ is obtained from the conservation law for the energy-momentum tensor as 
\begin{align}
 \frac{\dot{\bt}}{\bt} &= \frac{3\dot{A}\brkt{\frac{\pi^4}{12}+4Q_1+Q_2}+\dot{B}\brkt{2Q_2+Q_3}}{\frac{\pi^4}{4}+12Q_1+5Q_2+Q_3}, 
 \label{evolv:bt}
\end{align}
where $Q_3(x)$ is defined in \eqref{def:Q_3}. 
(See Appendix~\ref{conserv_law}.)
The profiles of $x^2Q_i(x)$ ($i=1,2,3$) are shown in Fig.~\ref{Q2}. 

We consider a situation that the moduli~$B$ and $\sgm$ have already been stabilized at $t=0$. 
Hence, we choose the initial conditions at $t=0$ as
\begin{align}
 A(0) &= 0, \;\;\;\;\;
 B(0) = B_*, \;\;\;\;\;
 \sgm(0) = \sgm_*, \;\;\;\;\;
 \bt(0) = \bt_{\rm I}, \nonumber\\
 \dot{A}(0) &= \sqrt{\frac{\hat{\rho}^{\rm tot}(0)}{3}}, \;\;\;\;\;
 \dot{B}(0) = \dot{\sgm}(0) = 0, 
 \label{initial_conds}
\end{align}
where $\bt_{\rm I}$ is a positive constant. 
The value of $\dot{A}(0)$ is determined by the constraint~\eqref{constraint}.

\section{Induced moduli oscillation and 3D scale factor}
\label{space_evolve}

\subsection{Moduli oscillation induced by $\mbox{\boldmath $p_2^{\rm rad}$}$}
As pointed out in our previous work~\cite{Otsuka:2022vgf}, the pressure for the compact space~$S^2$, $p_2^{\rm rad}$, pushes out 
the moduli from the potential minimum, 
and induces an oscillation of the moduli around the stabilized values in \eqref{static_sol}. 
Namely, even in the case that the moduli have been settled at the stabilized point before the radiation-dominated era, 
they will start to oscillate again. 
This effect cannot be neglected if the temperature is high enough compared to $m_{\rm KK}$. 

In order to see this behavior, we will see the time evolution of the moduli at early times. 
We assume that the radiation dominates the energy density at $t=0$, 
and the initial temperature is higher than $m_{\rm KK}$ (i.e., $\hat{\bt}_{\rm I}\equiv e^{-B_*}\bt_{\rm I}\ll 1$). 
Since we are interested in the oscillation around the stabilized values in \eqref{static_sol}, 
we define $\tl{B}\equiv B-B_*$ and $\tl{\sgm}\equiv \sgm-\sgm_*$. 
The mass eigenstates are linear combinations of them, which are defined as
\begin{align}
 \vph_1 &\equiv e^{\frac{3}{2}A}\brkt{2\cos\tht\tl{B}+\sin\tht\tl{\sgm}}, \nonumber\\
 \vph_2 &\equiv e^{\frac{3}{2}A}\brkt{-2\sin\tht\tl{B}+\cos\tht\tl{\sgm}}. 
\label{def:vph12}
\end{align}
The evolution equations for them are derived from \eqref{evolution_eqs} as
\begin{align}
 \begin{pmatrix} \ddot{\vph}_1 \\ \ddot{\vph}_2 \end{pmatrix} 
 &= -\begin{pmatrix} \lmd_1 & \\ & \lmd_2 \end{pmatrix}\begin{pmatrix} \vph_1 \\ \vph_2 \end{pmatrix}
 +2e^{\frac{3}{2}A}p_2^{\rm rad}\begin{pmatrix} \cos\tht \\ -\sin\tht \end{pmatrix}+\cdots, 
 \label{app:moduli_evolv}
\end{align}
where the ellipsis denotes higher order terms in $\vph_1$ or $\vph_2$, and 
\begin{align}
 \lmd_1 &\equiv 
 \frac{1}{2}\brkt{2m_{\rm KK}^2+m^2-\sqrt{4m_{\rm KK}^4+m^4}},  \nonumber\\
 \lmd_2 &\equiv 
 \frac{1}{2}\brkt{2m_{\rm KK}^2+m^2+\sqrt{4m_{\rm KK}^4+m^4}}, 
 \nonumber\\
 \tht &\equiv 
 \tan^{-1}\frac{2m_{\rm KK}^2}{m^2+\sqrt{4m_{\rm KK}^4+m^4}} = \tan^{-1}\frac{m_{\rm KK}^2}{\lmd_2-m_{\rm KK}^2}. 
 \label{def:lmd12}
\end{align}
We have neglected terms involving $\ddot{A}$, which are assumed to be small at initial times. 
When $m^2\gg m_{\rm KK}^2$, for example, these become $\lmd_1\simeq m_{\rm KK}^2$, 
$\lmd_2\simeq m^2$ and $\tht\simeq \lmd_1/\lmd_2$. 
 
In general, it is hard to solve \eqref{evolution_eqs} analytically because $A$, $B$ and $\sgm$ are coupled to each other. 
However, due to the assumption that the radiation is dominated at initial times, the expansion of the 3D space is determined 
only by $\rho^{\rm rad}$, and is almost independent of the moduli. 
Hence, we can treat the 3D expansion and the moduli oscillation separately. 
In fact, from \eqref{expr:rho-p} and \eqref{Qs:smallx}, 
the energy density and the pressure for the compact space~$S^2$ are approximately written as
\begin{align}
 \rho^{\rm rad} &\simeq \frac{10g_{\rm dof}}{\pi^3\bt^6}, \;\;\;\;\;
 p_2^{\rm rad} \simeq \frac{2g_{\rm dof}}{\pi^3\bt^6}, 
 \label{app:rhop2}
\end{align}
which are independent of the moduli. 
Notice that $\dot{A}\gg |\dot{B}|$ at the very early times 
because of the initial condition~\eqref{initial_conds}. 
Then, from \eqref{evolv:bt} and \eqref{Qs:smallx}, we obtain 
\begin{align}
 \frac{\dot{\bt}}{\bt} &\simeq \frac{3}{5}\dot{A}, 
\end{align}
which is immediately solved as $\bt\simeq \bt_{\rm I}e^{\frac{3}{5}A}$. 
Thus, \eqref{app:rhop2} is rewritten as
\begin{align}
 \rho^{\rm rad} &\simeq \Drad e^{-\frac{18}{5}A}, \;\;\;\;\;
 p_2^{\rm rad} \simeq \frac{\Drad}{5}e^{-\frac{18}{5}A}, 
\end{align}
where $\Drad\equiv 10g_{\rm dof}/(\pi^3\bt_{\rm I}^6)$. 
 
From \eqref{constraint}, we have 
\begin{align}
 \dot{A} &\simeq \sqrt{\frac{\rho^{\rm rad}}{3}} \simeq \sqrt{\frac{\Drad}{3}}e^{-\frac{9}{5}A}, 
 \label{dotA:earlytimes}
\end{align}
which leads to 
\begin{align}
 t \simeq \int_0^A dx\;\sqrt{\frac{3}{\Drad}}e^{\frac{9}{5}x} 
 = \frac{5}{3\sqrt{3\Drad}}\brkt{e^{\frac{9}{5}A}-1}. 
\end{align}
Therefore, we have 
\begin{align}
 e^A &\simeq \brkt{1+\sqrt{\CA}t}^{5/9}, \;\;\;\;\;
 \bt \simeq e^{\frac{3}{5}A} \simeq \brkt{1+\sqrt{\CA}t}^{1/3}, 
 \label{e^A:earlytimes}
\end{align}
where $\CA\equiv (27/25)\Drad=54g_{\rm dof}/(5\pi^3\bt_{\rm I}^6)$. 
This approximation is valid when $\hat{\bt} \equiv e^{-B_*}\bt<1$.  
This condition is translated as
\begin{align}
 t &< t_1 \equiv \frac{1}{\sqrt{\CA}\hat{\bt}_{\rm I}^3} = e^{3B_*}\sqrt{\frac{5\pi^3}{54g_{\rm dof}}}=0.038 e^{3B_*}. 
 \label{def:t1}
\end{align}
We have used that $\hat{\bt}_{\rm I}\ll 1$. 

Using \eqref{e^A:earlytimes}, \eqref{app:moduli_evolv} is rewritten as
\begin{align}
 \begin{pmatrix} \ddot{\vph}_1 \\ \ddot{\vph}_2 \end{pmatrix}
 &\simeq -\begin{pmatrix} \lmd_1 & \\ & \lmd_2 \end{pmatrix}
 \begin{pmatrix} \vph_1 \\ \vph_2 \end{pmatrix}
 +\frac{2\Drad}{5}\brkt{1+\CA t}^{-7/6}\begin{pmatrix} \cos\tht \\ -\sin\tht \end{pmatrix}. 
 \label{moduli_evolve_eqs:app}
\end{align}
From \eqref{initial_conds}, the initial conditions at $t=0$ are read off as
\begin{align}
 \vph_1(0) &= \vph_2(0) = \dot{\vph}_1(0) = \dot{\vph}_2(0) = 0. 
\end{align}
The solution is expressed as
\begin{align}
 \vph_1(t) &\simeq -\frac{2\Drad\cos\tht}{5\lmd_1}\brkt{\frac{\lmd_1}{\CA}}^{\frac{7}{12}}
 \Im\brc{e^{\frac{\pi}{12}i}\cU_{7/6}(t;\lmd_1,\CA)} 
 \nonumber\\
 &= -\frac{10\cos\tht}{27}\brkt{\frac{\CA}{\lmd_1}}^{\frac{5}{12}}\Im\brc{e^{\frac{\pi}{12}i}\cU_{7/6}(t;\lmd_1,\CA)}, \nonumber\\
 \vph_2(t) &\simeq \frac{2\Drad\sin\tht}{5\lmd_2}\brkt{\frac{\lmd_2}{\CA}}^{\frac{7}{12}}
 \Im\brc{e^{\frac{\pi}{12}i}\cU_{7/6}(t;\lmd_2,\CA)} 
 \nonumber\\
 &= \frac{10\sin\tht}{27}\brkt{\frac{\CA}{\lmd_2}}^{\frac{5}{12}}\Im\brc{e^{\frac{\pi}{12}i}\cU_{7/6}(t;\lmd_2,\CA)}. 
 \label{vphs:earlytimes}
\end{align}
(See Appendix~\ref{forced_osc}.)
The function~$\cU_q(t;\lmd,C)$ is defined by the incomplete gamma function as \eqref{def:cU}. 
From these and \eqref{e^A:earlytimes}, 
we obtain an approximate solution of the moduli evolution equations at early times. 
We have checked that this approximate solution agrees well with the solution of the full evolution equation~\eqref{evolution_eqs} 
obtained by the numerical computation. 

\begin{figure}[t]
  \begin{center}
    \includegraphics[scale=0.6]{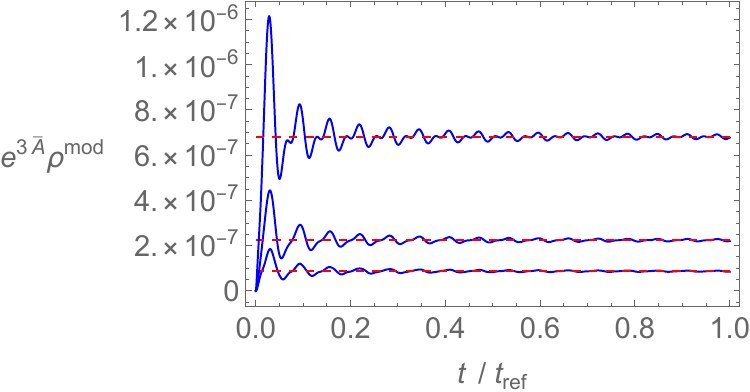} \;\;\;\;\;
    \includegraphics[scale=0.6]{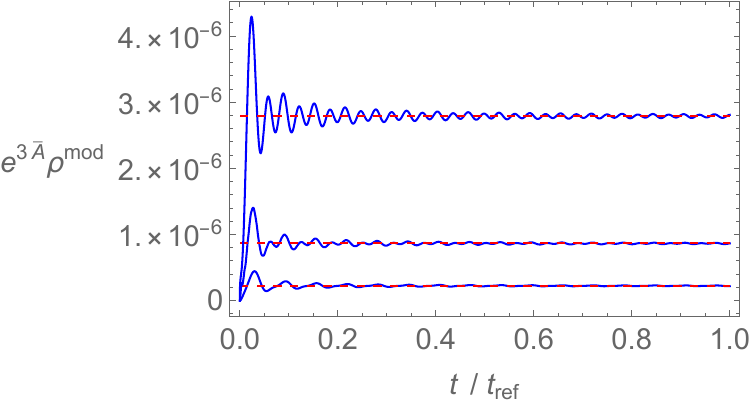} 
  \end{center}
\caption{The profiles of $e^{3\bar{A}}\rho^{\rm mod}$ as a function of $t/\tref$, 
where $\tref\equiv 100/\sqrt{\lmd_1}$. 
In the left plot, we choose $\sgm_*=10,12,14$ from bottom to top with $m=10^{-2}$ and $\bt_{\rm I}=20$. 
In the right plot, we choose $m=10^{-2}$, $10^{-3}$, $10^{-4}$ from bottom to top 
with $\sgm_*=12$ and $\bt_{\rm I}=20$.
The dashed lines denote the asymptotic values given by \eqref{def:Cmod}.}
    \label{profile:rho^mod}
\end{figure}

As we will see in the next subsection, it is convenient to define
\begin{align}
 \bar{A}(t) &\equiv A(t)+\tl{B}(t).  
 \label{def:bA}
\end{align}
Since $|\tl{B}(t)|\ll A(t)$ except for an early short period $0\leq t\ll\CA^{-1/2}$, 
$e^{\bar{A}}$ can be understood as a modified 3D scale factor. 
The mixing term between $\dot{A}$ and $\dot{B}$ in the constraint~\eqref{constraint} is 
absorbed into $\dot{\bar{A}}^2$, and \eqref{constraint} is rewritten in a similar form to the 4D Friedmann equation.\footnote{
Note that $\dot{\bar{A}}$ corresponds to the Hubble expansion rate. 
}
\begin{align}
 3\dot{\bar{A}}^2 &= \rho^{\rm mod}+\rho^{\rm rad}, 
 \label{3barA^2}
\end{align}
where
\begin{align}
 \rho^{\rm mod} &\equiv 2\dot{B}^2+\frac{1}{2}\dot{\sgm}^2
 +2\brkt{e^{-\frac{\sgm}{2}}-\frac{e^{\frac{\sgm}{2}-2B}}{4}}^2+\frac{m^2}{2}\brkt{\sgm-\sgm_*}^2, 
\end{align}
represents the energy density of the moduli oscillation. 
The moduli energy density~$\rho^{\rm mod}$ is expressed in terms of $\vph_1$ and $\vph_2$ as 
\begin{align}
 \rho^{\rm mod} &= \frac{e^{-3\bar{A}}}{2}\brc{\brkt{\dot{\vph}_1-\frac{3}{2}\dot{\bar{A}}\vph_1}^2
 +\brkt{\dot{\vph}_2-\frac{3}{2}\dot{\bar{A}}\vph_2}^2+\lmd_1\vph_1^2+\lmd_2\vph_2^2}+\cdots, 
 \label{app:rho^mod}
\end{align}
where the ellipsis denotes higher order terms in $\vph_1$ or $\vph_2$. 
Using \ref{vphs:earlytimes}, we can plot $e^{3\bar{A}}\rho^{\rm mod}(t)$ as Fig.~\ref{profile:rho^mod}. 
From this plot, we can see that $e^{3\bar{A}}\rho^{\rm mod}$ is almost a constant for $t>\tref$, 
where 
\begin{align}
 \tref &\equiv \frac{100}{\sqrt{\lmd_1}} \sim 100\max\brkt{\frac{\sqrt{2}}{m},\frac{1}{m_{\rm KK}}}. 
\end{align}

Noting that $\abs{\Gm(1-q,ix)}\simeq x^{-q}$ for $x\gg 1$, we find that 
\begin{align}
 \lim_{t\to\infty}e^{3\bar{A}(t)}\rho^{\rm mod}(t) 
 &= \frac{50\CA^{5/6}}{729}\brkt{\lmd_1^{1/6}\cos^2\tht\abs{\Gm\brkt{-\frac{1}{6},i\sqrt{\frac{\lmd_1}{\CA}}}}^2
 +\lmd_2^{1/6}\sin^2\tht\abs{\Gm\brkt{-\frac{1}{6},\sqrt{\frac{\lmd_2}{\CA}}}}^2} \nonumber\\
 &\equiv \Cmod. 
 \label{def:Cmod}
\end{align}
We have used \eqref{dotvph}. 
Namely, $\rho^{\rm mod}$ decays as 
\begin{align}
 \rho^{\rm mod}(t) &\simeq \Cmod e^{-3\bar{A}(t)}. 
 \label{asymp:rho^mod}
\end{align}
at late times ($t>\tref$). 

\begin{figure}[t]
  \begin{center}
    \includegraphics[scale=0.60]{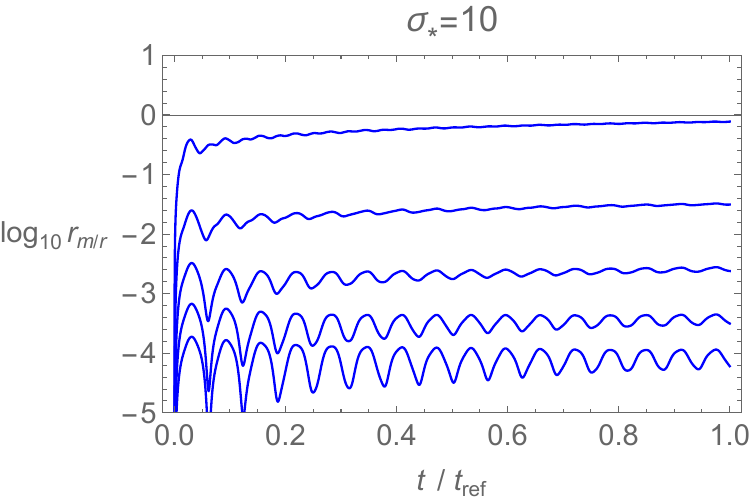} \;\;\;\;\;
    \includegraphics[scale=0.60]{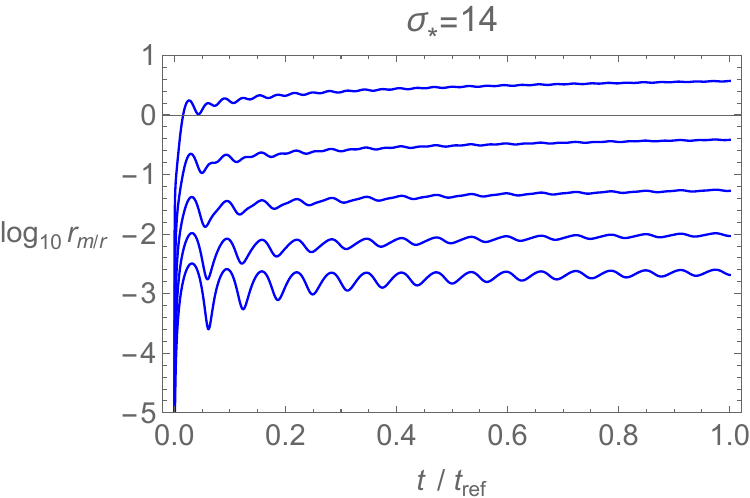} 
  \end{center}
\caption{The logarithm of the function~$r_{\rm m/r}$ in \eqref{def:rmr} as a function of $t/\tref$. 
The lines correspond $\bt_{\rm I}=10,20,30,40,50$ from top to bottom. 
The other parameters are chosen as $m=0.01$, $\sgm_*=10$ (left plot) and $\sgm_*=14$ (right plot). 
}
    \label{profile:rmr}
\end{figure}

Here, we comment on the validity of the above approximations. 
We have used that $\rho^{\rm rad}\gg \rho^{\rm mod}$ to obtain \eqref{e^A:earlytimes}. 
When this condition is satisfied, the ratio of $\rho^{\rm mod}$ to $\rho^{\rm rad}$ is 
\begin{align}
 \frac{\rho^{\rm mod}}{\rho^{\rm rad}} 
 &\simeq \frac{27}{25\CA}\brc{\brkt{\dot{\vph}_1-\frac{3}{2}\dot{\bar{A}}\vph_1}^2
 +\brkt{\dot{\vph}_2-\frac{3}{2}\dot{\bar{A}}\vph_2}^2+\lmd_1\vph_1^2+\lmd_2\vph_2^2}
 \brkt{1+\sqrt{\CA}t}^{\frac{1}{3}} \nonumber\\
 &\equiv r_{\rm m/r}(t), 
 \label{def:rmr}
\end{align}
where $\vph_1(t)$ and $\vph_2(t)$ are given by \eqref{vphs:earlytimes}, 
and $\dot{\bar{A}}(t)=5\sqrt{\CA}/\brc{9\brkt{1+\sqrt{\CA}t}}$. 
Note that the function~$r_{\rm m/r}(t)$ is determined when the model parameters~${m,\sgm_*}$ 
and the initial condition~$\bt_{\rm I}$ are given. 
This ratio is plotted in Fig.~\ref{profile:rmr}. 
We can see from the plots that the approximate solution in \eqref{vphs:earlytimes} is no longer valid when $t\simeq \tref$ 
in the case of $\bt_{\rm I}=10$, $m=0.01$ and $\sgm_*=14$. 
As a general property, the ratio~$\rho^{\rm mod}/\rho^{\rm rad}$ takes a larger value  
for higher initial temperature or for shallower moduli potential (i.e., smaller $m$ or larger $\sgm_*$). 
When $\rho^{\rm mod}$ approaches to $\rho^{\rm rad}$, the expansion rate for the 3D space 
becomes larger than \eqref{dotA:earlytimes}, 
and the inhomogeneous term in \eqref{app:moduli_evolv} decays more rapidly than \eqref{moduli_evolve_eqs:app}. 
After the inhomogeneous term becomes negligible, the solution will reduce to a linear combination of 
two simple harmonic oscillators, and $e^{3A}\rho^{\rm mod}$ becomes constant. 
Thus, the constant $\Cmod$ in \eqref{asymp:rho^mod} takes a smaller value than \eqref{def:Cmod} 
if $r_{\rm m/r}$ is close to one before $t=\tref$.

\subsection{Smoothed 3D scale factor}

\begin{figure}[t]
  \begin{center}
    \includegraphics[scale=0.57]{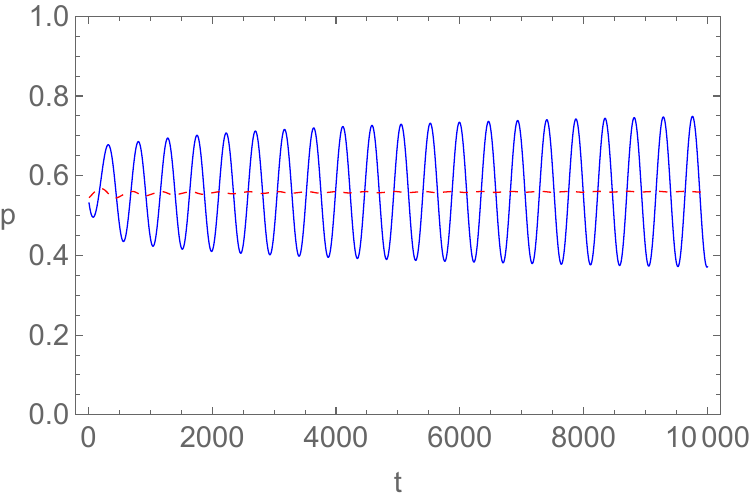} 
  \end{center}
\caption{The profiles of $(t-\tc)\dot{A}(t)$ (solid line) and $p(t)$ defined in \eqref{def:p} (dashed line) 
in the case of $m=0.1$, $\sgm_*=10$, $\bt_{\rm I}=10$, and $\tc=-125$ in the unit of $M_6$. }
    \label{profile:p-t}
\end{figure}

In the usual 4D cosmology, the 3D scale factor evolves as $e^{A(t)}\propto t^{1/2}$ in the radiation-dominated era 
and as $e^{A(t)}\propto t^{2/3}$ in the matter-dominated era. 
Thus, it is convenient to define the effective power~$p$ as $p\equiv t\dot{A}$. 
Then, $p=1/2$ (2/3) in the radiation- (matter-) dominated era.  
However, as we pointed out in Ref.~\cite{Otsuka:2022vgf}, this quantity oscillates due to the effect of the moduli oscillation 
(see Fig.~\ref{profile:p-t}). 
Thus, we modify the definition of $p$ as
\begin{align}
 p &\equiv (t-\tc)\dot{\bar{A}}, 
 \label{def:p}
\end{align}
where $\bar{A}$ is defined by \eqref{def:bA}. 
The constant~$\tc$ is chosen so that $p$ is almost independent of $t$ at early times. 
We will show how to choose $t_{\rm c}$ in Sec.~\ref{choice_tc}. 
As we can see from Fig.~\ref{profile:p-t}, this modification removes the effect of the moduli oscillation~\cite{Otsuka:2022vgf}. 

\begin{figure}[t]
  \begin{center}
    \includegraphics[scale=0.60]{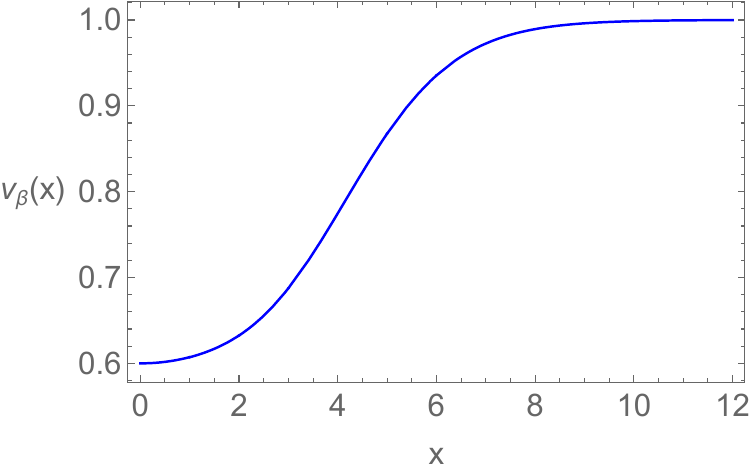} \hspace{5mm}
    \includegraphics[scale=0.60]{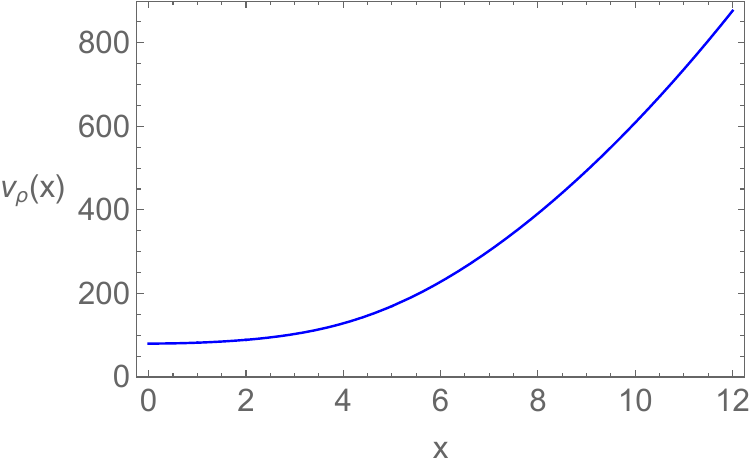} 
  \end{center}
\caption{The profiles of $v_\bt(x)$ and $v_\rho(x)$. }
    \label{profile:vbtrho}
\end{figure}

For $t\geq\tref$, the radiation energy density in \eqref{expr:rho-p:text} is approximated as
\begin{align}
 \rho^{\rm rad} &\simeq \Crad\frac{v_\rho(\hat{\bt})}{\hat{\bt}^6}, \nonumber\\
  \Crad &\equiv \frac{g_{\rm dof}e^{-6B_*}}{8\pi^3}, 
 \label{fct:rho^rad-hatbt}
\end{align}
where $\hat{\bt}\equiv e^{-B_*}\bt=m_{\rm KK}\bt$ is the inverse temperature in the unit of the KK mass~$m_{\rm KK}$, and 
\begin{align}
 v_\rho(x) &\equiv x^2\brc{\frac{\pi^4}{16}+3Q_1(x)+Q_2(x)}. 
 \label{def:v_rho}
\end{align}
The evolution equation for $\bt$~\eqref{evolv:bt} is approximated as
\begin{align}
 \frac{\dot{\bt}}{\bt} &\simeq v_\bt(\hat{\bt})\dot{\bar{A}}, 
 \label{app:bt-A}
\end{align}
where 
\begin{align}
 v_\bt(x) &\equiv \frac{\frac{\pi^4}{4}+12Q_1(x)+3Q_2(x)}{\frac{\pi^4}{4}+12Q_1(x)+5Q_2(x)+Q_3(x)}. 
 \label{def:v_bt}
\end{align}
The profiles of $v_\bt(x)$ and $v_\rho(x)$ are shown in Fig.~\ref{profile:vbtrho}

\begin{figure}[t]
  \begin{center}
    \includegraphics[scale=0.55]{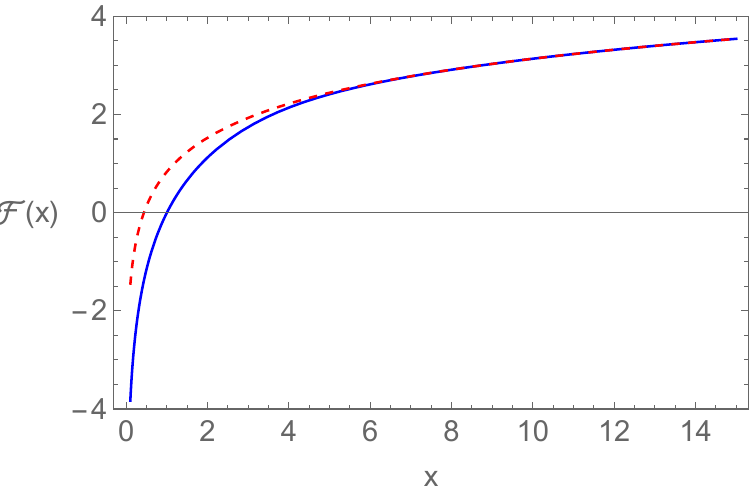} 
  \end{center}
\caption{The profile of $\cF(x)$. The dotted line represents $\ln x+0.833$. }
    \label{profile:F}
\end{figure}

Solving \eqref{app:bt-A}, the (smoothed) 3D scale factor~$\bar{A}$ is expressed as a function of $\hat{\bt}$, 
\begin{align}
 \bar{A}(\hat{\bt}) &\simeq \bAref+\int_{\hat{\bt}_{\rm ref}}^{\hat{\bt}}\frac{dx}{xv_\bt(x)} \nonumber\\
 &= \bAref+\cF(\hat{\bt})-\cF(\btref), 
 \label{fct:A-hatbt}
\end{align}
where $\bAref$ and $\btref$ denote the values at $t=\tref$, and
\begin{align}
 \cF(x) &\equiv \int_1^x\frac{dy}{yv_\bt(y)}. 
\end{align}
The profile of $\cF(x)$ is shown in Fig.~\ref{profile:F}. 

\subsection{Expression of effective power~$\mbox{\boldmath $p$}$}
\begin{figure}[t]
  \begin{center}
    \includegraphics[scale=0.60]{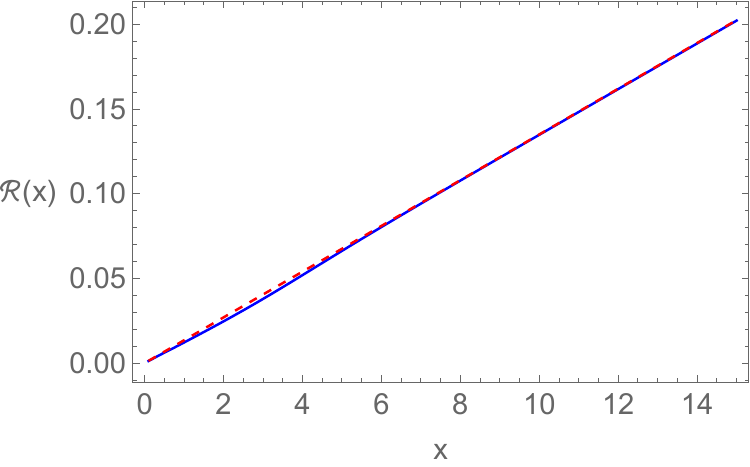}
  \end{center}
\caption{The profile of $\cR(x)$. The dotted represents $0.0135x$. }
    \label{profile:R}
\end{figure}

Here, we derive an explicit expression for the effective power~$p$. 
From \eqref{3barA^2} and \eqref{asymp:rho^mod}, we have~\footnote{
We focus on the positive solution of $\dot{\bar{A}}$ since we are interested in an expanding 3D universe. 
}
\begin{align}
 \dot{\bar{A}} &\simeq \sqrt{\frac{1}{3}\brkt{\Cmod e^{-3\bar{A}}+\rho^{\rm rad}}}, 
 \label{dotbarA}
\end{align}
which leads to
\begin{align}
 t-\tref &\simeq \int_{\hat{\bt}_{\rm ref}}^{\hat{\bt}} dx\;\frac{d\bar{A}}{d\hat{\bt}}(x)
 \sqrt{\frac{3}{\Cmod e^{-3\bar{A}(x)}+\rho^{\rm rad}(x)}} \nonumber\\
 &= \int_{\btref}^{\hat{\bt}}\frac{dx}{xv_\bt(x)}\;\sqrt{\frac{3}{\Cmod e^{-3\bar{A}(x)}+\Crad x^{-6} v_\rho(x)}} \nonumber\\
 &= \sqrt{\frac{3}{\Crad}}\int_{\btref}^{\hat{\bt}}dx\;\frac{x^2}{v_\bt(x)\sqrt{v_\rho(x)}}\brc{1+\Rmr\cR(x)}^{-1/2}, 
\end{align}
where 
\begin{align}
 \Rmr &\equiv \frac{\Cmod e^{-3\bAref+3\cF(\btref)}}{\Crad}, \;\;\;\;\;
 \cR(x) \equiv \frac{x^6 e^{-3\cF(x)}}{v_\rho(x)}. 
 \label{def:cR}
\end{align}
We have used \eqref{fct:A-hatbt} at the last step. 
As shown in Fig.~\ref{profile:R}, $\cR(x)$ is well approximated as a linear function. 
\begin{align}
 \cR(x) &\simeq 0.0135x. 
 \label{ap:calR}
\end{align}
Thus, the above expression can be approximated as
\begin{align}
 t &\simeq \tref+\sqrt{\frac{3}{\Crad}}\int_{\btref}^{\hat{\bt}}dx\;\frac{x^2}{v_\bt(x)\sqrt{v_\rho(x)}}
 \brkt{1+\frac{x}{\btmod}}^{-1/2}, 
 \label{t-hatbt}
\end{align}
where
\begin{align}
 \btmod &\equiv \frac{1}{0.0135\Rmr} \simeq \frac{74.1}{\Rmr}. 
\end{align}

Using this expression, the effective power~$p$ defined in \eqref{def:p} is expressed as
\begin{align}
 p(\hat{\bt}) &= (t-\tc)\frac{d\bar{A}}{d\hat{\bt}}(\hat{\bt})\brc{\frac{dt}{d\hat{\bt}}(\hat{\bt})}^{-1} \nonumber\\
 &\simeq \brc{\tref-\tc+\int_{\hat{\bt}_{\rm ref}}^{\hat{\bt}}\frac{dx}{xv_\bt(x)}\;
 \sqrt{\frac{3}{\Cmod e^{-3\bar{A}(x)}+\Crad x^{-6}v_\rho(x)}}} \nonumber\\
 &\quad
 \times \sqrt{\frac{\Cmod e^{-3\bar{A}(\hat{\bt})}+\Crad \hat{\bt}^{-6}v_\rho(\hat{\bt})}{3}} \nonumber\\
 &\simeq \brc{\tref-\tc+\sqrt{\frac{3}{\Crad}}\int_{\btref}^{\hat{\bt}}dx\;\frac{x^2}{v_\bt(x)\sqrt{v_\rho(x)}}
 \brkt{1+\frac{x}{\btmod}}^{-1/2}} \nonumber\\
 &\quad
 \times \sqrt{\frac{\Crad v_\rho(\hat{\bt})}{3\hat{\bt}^6}}\brkt{1+\frac{\hat{\bt}}{\btmod}}^{1/2}. 
 \label{expr:p}
\end{align}

\subsection{Choice of $\mbox{\boldmath $\tc$}$}
\label{choice_tc}
As we mentioned, the constant~$\tc$ is determined so that $p$ is almost independent of $t$ at early times~$t\leq\tref$. 
By assumption, $\rho^{\rm mod}(\tref)\simeq\Cmod e^{-3\bAref}\ll \rho^{\rm rad}(\tref)\simeq\Crad \btref^{-6}v_\rho(\btref)$, 
i.e., $\btref\ll\btmod$. 
Then, when $t\sim\tref$, \eqref{t-hatbt} is approximated as
\begin{align}
 t(\hat{\bt}) &= \tref+\sqrt{\frac{3}{\Crad}}\int_{\hat{\bt}_{\rm ref}}^{\hat{\bt}}dx\;\frac{x^2}{v_\bt(x)\sqrt{v_\rho(x)}}
 \brkt{1-\frac{x}{2\btmod}+\cdots} \nonumber\\
 &\simeq \tref+\sqrt{\frac{3}{\Crad}}\sbk{\cG(x)-\frac{1}{2\btmod}\cH(x)}_{\hat{\bt}_{\rm ref}}^{\hat{\bt}}, 
\end{align}
where
\begin{align}
 \cG(x) &\equiv \int_0^x dy\;\frac{y^2}{v_\bt(y)\sqrt{v_\rho(y)}}, \nonumber\\
 \cH(x) &\equiv \int_0^x dy\;\frac{y^3}{v_\bt(y)\sqrt{v_\rho(y)}}. 
 \label{def:GH}
\end{align}
\begin{figure}[t]
  \begin{center}
    \includegraphics[scale=0.58]{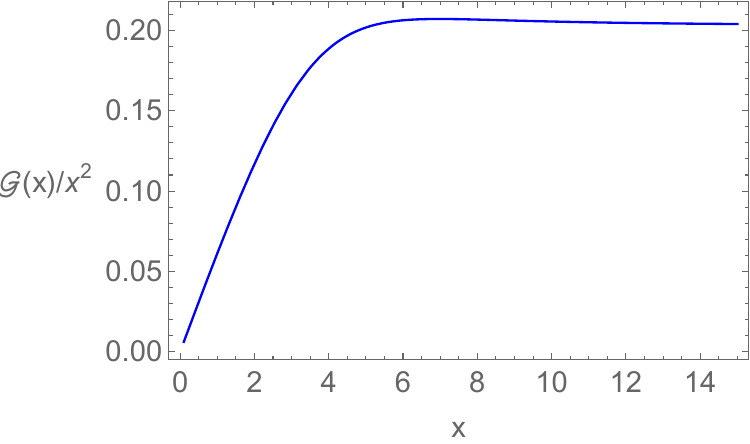} \hspace{5mm}
    \includegraphics[scale=0.60]{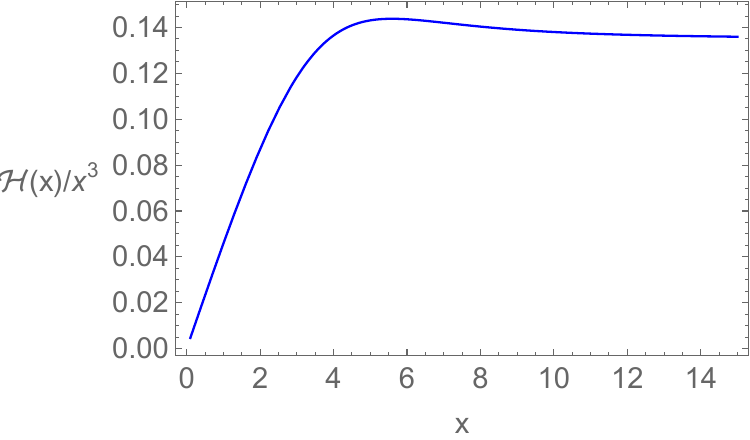}
  \end{center}
\caption{The profiles of $\cF(x)$, $\cG(x)/x^2$ and $\cH(x)/x^3$. }
    \label{profile:FGH}
\end{figure}
Fig.~\ref{profile:FGH} shows the profiles of $\cG(x)/x^2$ and $\cH(x)/x^3$.

Thus, when $t\sim \tref$, \eqref{expr:p} is approximated as
\begin{align}
 p(\hat{\bt}) 
 &\simeq \brc{\tref-\tc+\sqrt{\frac{3}{\Crad}}\sbk{\cG(x)-\frac{1}{2\btmod} \cH(x)}_{\hat{\bt}_{\rm ref}}^{\hat{\bt}}} 
 \sqrt{\frac{\Crad v_\rho(\hat{\bt})}{3\hat{\bt}^6}}\brkt{1+\frac{\hat{\bt}}{2\btmod}}. 
 \label{expr:p:2}
\end{align}
Here, we choose $\tc$ as 
\begin{align}
 \tc &= \tref-\sqrt{\frac{3}{\Crad}}\brc{\cG(\hat{\bt}_{\rm ref})-\frac{1}{2\btmod}\cH(\hat{\bt}_{\rm ref})}. 
 \label{choice:t_0}
\end{align}
Then, \eqref{expr:p:2} becomes
\begin{align}
 p(\hat{\bt}) &\simeq \brc{\cG(\hat{\bt})-\frac{1}{2\btmod}\cH(\hat{\bt})}\frac{\sqrt{v_\rho(\hat{\bt})}}{\hat{\bt}^3}
 \brkt{1+\frac{\hat{\bt}}{2\btmod}}. 
\end{align}
For $\hat{\bt}\ll 1$, the functions we have defined behave as
\begin{align}
 v_\bt(\hat{\bt}) &\simeq \frac{3}{5}, \;\;\;\;\;
 v_\rho(\hat{\bt}) \simeq 80, \nonumber\\
 \cF(\hat{\bt}) &\simeq \frac{5}{3}\ln\hat{\bt}, \;\;\;\;\;
 \cG(\hat{\bt}) \simeq \frac{\sqrt{5}}{36}\hat{\bt}^3, \;\;\;\;\;
 \cH(\hat{\bt}) \simeq \frac{\sqrt{5}}{48}\hat{\bt}^4. 
\end{align}
By assumption, $\hat{\bt}\ll\btmod$. 
Thus, $p(\hat{\bt})$ behaves as~\footnote{
This value corresponds to the 6D radiation-dominated universe, 
which is also obtained from $e^A\propto t^{2/(3(1+w))}$ with $w=1/5$. 
Note that $w^{-1}$ measures the space dimensions that the radiation feels. 
}
\begin{align}
 p(\hat{\bt}) &\simeq \cG(\hat{\bt})\frac{\sqrt{v_\rho(\hat{\bt})}}{\hat{\bt}^3} \simeq \frac{5}{9}, 
\end{align}
which is independent of $\hat{\bt}$ (or $t$). 
Hence, the choice of $t_0$ in \eqref{choice:t_0} is appropriate. 

Using this choice of $t_0$, \eqref{expr:p} becomes
\begin{align}
 p(\hat{\bt}) &\simeq \brc{\int_{\btref}^{\hat{\bt}}dx\;\frac{x^2}{v_\bt(x)\sqrt{v_\rho(x)}}\brkt{1+\frac{x}{\btmod}}^{-1/2}
 +\cG(\btref)-\frac{\cH(\btref)}{2\btmod}} 
 \nonumber\\
 &\quad
 \times \frac{\sqrt{v_\rho(\hat{\bt})}}{\hat{\bt}^3}\brkt{1+\frac{\hat{\bt}}{\btmod}}^{1/2}. 
 \label{p-hbt}
\end{align}
Combining this with \eqref{t-hatbt}, we can plot $p$ as a function of $t$.

\section{Time evolution of 3D space}
\label{evolve:p}
In this section, we discuss the expansion of the 3D space by evaluating the time evolution of the effective power~$p$. 

\subsection{Parameter dependence of effective power~$\mbox{\boldmath $p$}$}
When $\hat{\bt}\ll \btmod$, \eqref{t-hatbt} and \eqref{p-hbt} reduce to
\begin{align}
 t(\hat{\bt}) &\simeq \tref+\sqrt{\frac{3}{\Crad}}\brc{\cG(\hat{\bt})-\cG(\btref)}, \nonumber\\
 p(\hat{\bt}) &\simeq \cG(\hat{\bt})\frac{\sqrt{v_\rho(\hat{\bt})}}{\hat{\bt}^3}. 
 \label{p:smallhatbt}
\end{align}
\begin{figure}[t]
  \begin{center}
    \includegraphics[scale=0.60]{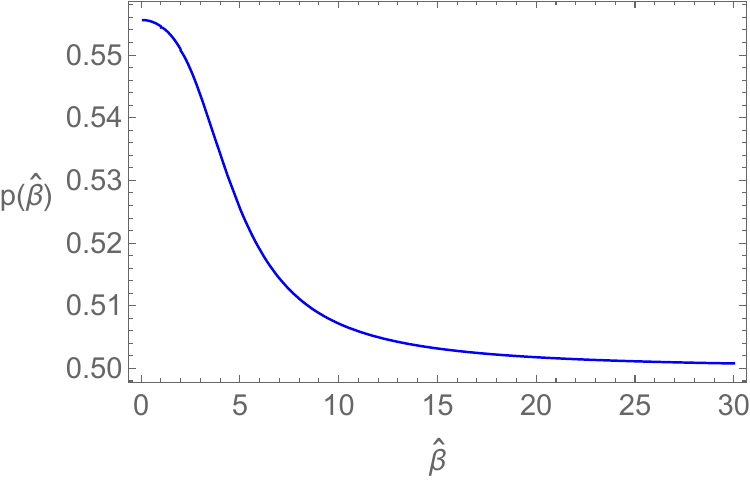}
  \end{center}
\caption{The profile of the function in \eqref{p:smallhatbt}. }
    \label{p:smallbeta}
\end{figure}
From Fig.~\ref{p:smallbeta}, we can see that the power~$p$ changes its value from 9/5 to 1/2 during the period~$2<\hat{\bt}<10$. 
If $\btmod\gg 15$, the 3D space expands as in the 4D radiation-dominated era until $\hat{\bt}$ approaches to $\btmod$. 

When $\hat{\bt}\gg \btmod$, on the other hand, the contribution of the moduli oscillation dominates the energy density, 
and $p(\hat{\bt})$ in \eqref{p-hbt} can be estimated as
\begin{align}
 t(\hat{\bt}) &\simeq \tref+\sqrt{\frac{3}{\Crad}}\int_{\btref}^{\hat{\bt}}dx\;\frac{x^2}{v_\bt(x)\sqrt{v_\rho(x)}}\sqrt{\frac{\btmod}{x}} \nonumber\\
 &\simeq \tref+\sqrt{\frac{3\btmod}{\Crad}}\int_0^{\hat{\bt}}dx\;\frac{x^{3/2}}{\frac{\pi^2}{4}x} 
 = \tref+\frac{8}{3\pi^2}\sqrt{\frac{3\btmod}{\Crad}}\hat{\bt}^{3/2}, 
 \nonumber\\
 p(\hat{\bt}) &\simeq \brc{\int_0^{\hat{\bt}}dx\;\frac{x^{3/2}}{v_\bt(x)\sqrt{v_\rho(x)}}}
 \frac{\sqrt{v_\rho(\hat{\bt})}}{\hat{\bt}^{5/2}} \nonumber\\
 &\simeq \brc{\int_0^{\hat{\bt}}dx\;\frac{x^{3/2}}{\frac{\pi^2}{4}x}}\frac{\frac{\pi^2}{4}\hat{\bt}}{\hat{\bt}^{5/2}} = \frac{2}{3}, 
\end{align}
where we have used that
\begin{align}
 v_\bt(x) &\simeq 1, \;\;\;\;\;
 v_\rho(x) \simeq \frac{\pi^4}{16}x^2, 
\end{align}
for $x\gg 1$. 

\begin{figure}[t]
  \begin{center}
    \includegraphics[scale=0.60]{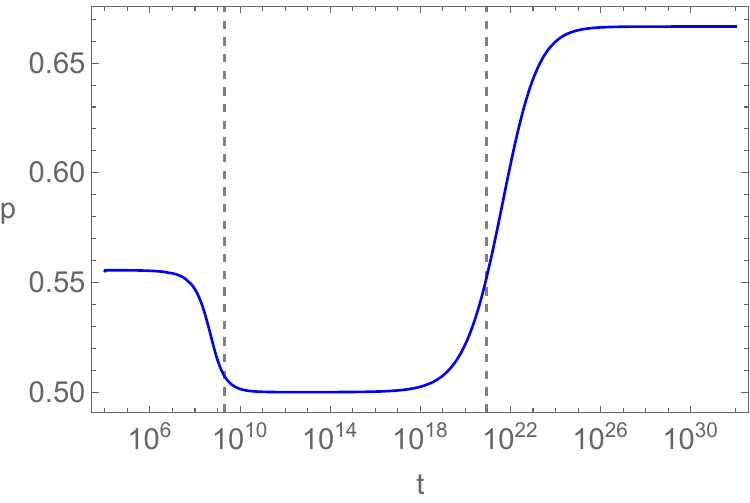}
  \end{center}
\caption{The effective power~$p$ as a function of $t$. 
The parameters are chosen as $\sgm_*=14$, $\btref=10^{-2}$ and $\Rmr=10^{-5}$. 
The left and right vertical dashed lines denote $t=\trad$ and $t=\tmod$, respectively. }
    \label{profile:pw}
\end{figure}

Thus, if we define
\begin{align}
 \trad &\equiv t(\hat{\bt}=10), \;\;\;\;\;
 \tmod \equiv t(\btmod), 
 \label{def:t_radmod}
\end{align}
the power~$p$ changes from 9/5 to 1/2 around $t=\trad$, and from 1/2 to 2/3 around $t=\tmod$. 
We show a typical profile of the function~$p(t)$ in Fig.~\ref{profile:pw}. 
The parameters are chosen as $\sgm_*=14$, $\btref=10^{-2}$ and $\Rmr=10^{-5}$. 

\begin{figure}[t]
  \begin{center}
    \includegraphics[scale=0.60]{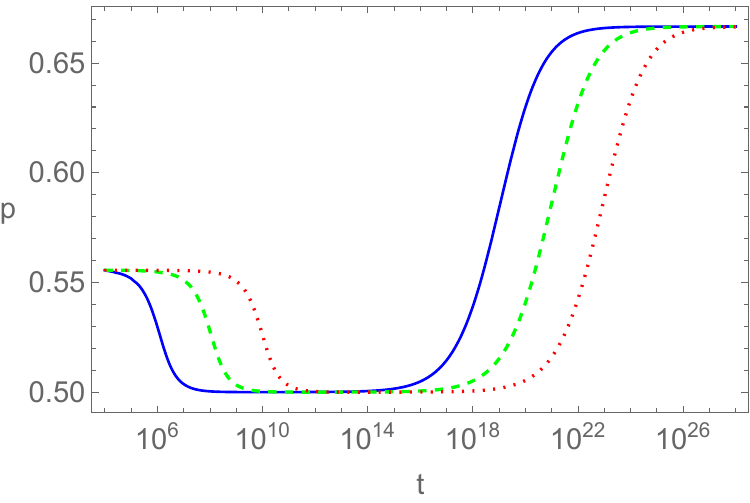} 
  \end{center}
\caption{The profile of $p(t)$ for various values of $\sgm_*$. 
The solid, dashed, and dotted lines correspond to $\sgm_*=10$, 13 and 16, respectively. 
The other parameters are chosen as $\tref=10^4$, $\btref=10^{-2}$ and $\Rmr=10^{-5}$. }
    \label{profile:pw4}
\end{figure}

\begin{figure}[t]
  \begin{center}
    \includegraphics[scale=0.55]{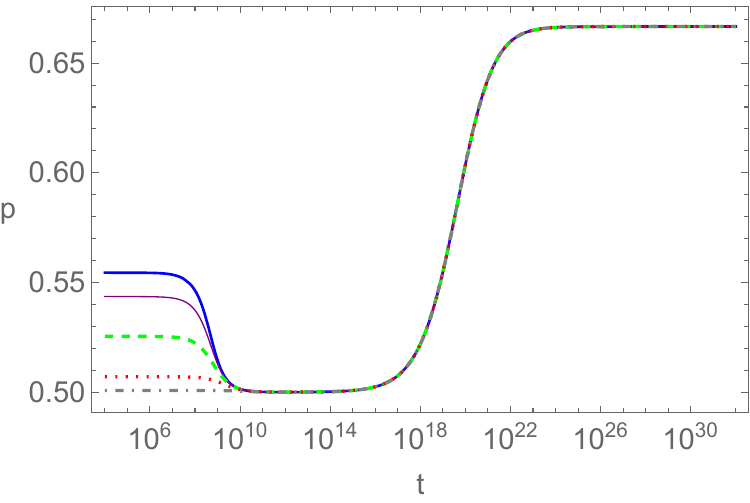} \;\;\;\;\;
    \includegraphics[scale=0.55]{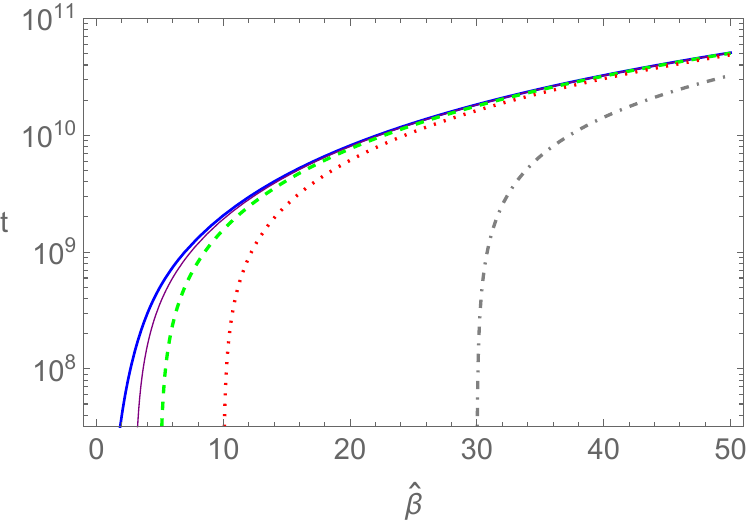}
  \end{center}
\caption{The profiles of $p(t)$ (left) and $t(\hat{\bt})$ (right). 
The thick, thin, dashed, dotted and dotdashed lines correspond to $\btref=1$, 3, 5, 10 and 30, respectively. 
The other parameters are chosen as $\tref=10^4$, $\sgm_*=14$ (i.e., $\Crad=2.97 \times 10^{-16}$), 
and $\Rmr=10^{-5}$. 
}
    \label{profile:t-hatbt}
\end{figure}

\begin{figure}[t]
  \begin{center}
    \includegraphics[scale=0.55]{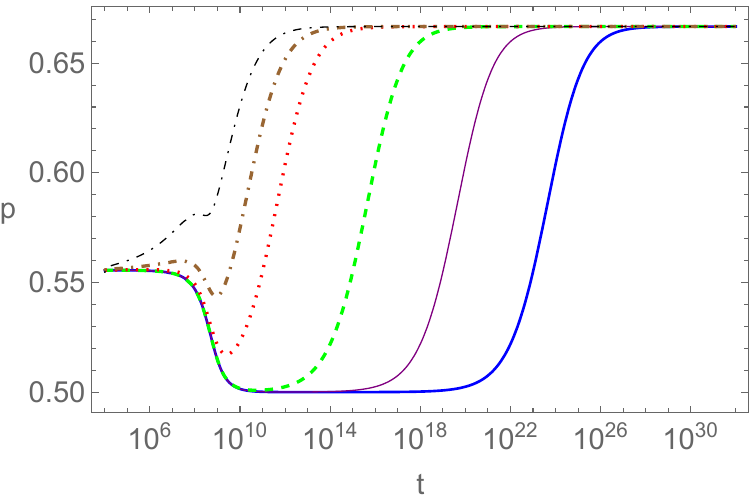} 
  \end{center}
\caption{The profile of $p(t)$. 
The thick, thin, dashed, dotted, thick dotdashed and thin dotdashed lines correspond to $\Rmr=10^{-6}$, $10^{-4}$, $10^{-2}$, 1, 5 and 20, respectively. 
The other parameters are chosen as $\tref=10^4$, $\btref=0.01$ and $\sgm_*=14$ (i.e., $\Crad=2.97 \times 10^{-16}$). 
}
    \label{profile:pw3}
\end{figure}

As we can see from \eqref{t-hatbt} and \eqref{p-hbt}, the function~$p(t)$ depends on the parameters 
only through $\tref$, $\Crad$, $\btref$ and $\Rmr$. 
Among them, we choose $\tref$ much smaller than the second term of $t(\hat{\bt})$ in \eqref{t-hatbt}, and thus 
its dependence can be neglected. 
Let us see the dependences on the other three parameters individually. 
\begin{description}
\item[\mbox{\boldmath $\Crad$}-dependence] \mbox{}\\
From \eqref{fct:rho^rad-hatbt}, $\Crad$ is determined only by $\sgm_*$ (or $m_{\rm KK}$), 
and it only affects the overall time scale if $\tref$ is negligible (see \eqref{t-hatbt}). 
Fig.~\ref{profile:pw4} shows the profile of $p(t)$ for different values of $\sgm_*$. 
The solid, dashed and dotted lines correspond $\sgm_*=10$, 13 and 16, respectively. 
As this plot shows, the value of $\sgm_*$ (i.e., $\Crad$) just shifts the profile to the time direction without changing its shape.

\item[\mbox{\boldmath $\btref$}-dependence] \mbox{}\\
The left plot in Fig.~\ref{profile:t-hatbt} shows the profile of $p(t)$ for various values of $\btref$. 
For $\btref\lesssim 1$, the profile of $p(t)$ is almost independent of $\btref$, 
and the initial value of $p(t)$ for $t\ll \trad$ is 9/5, which is the value of the 6D radiation-dominated universe. 
For $2<\btref<20$, the value of $p(t)$ at early times ($t\ll\trad$) decreases as $\btref$ increases. 
This can be understood from Fig.~\ref{p:smallbeta}. 
A larger value of $\btref$ in this region indicates that the temperature is not high enough for the radiation to feel the compact space~$S^2$ completely, 
and the 3D space expands less rapidly. 
For $\btref>15$, the radiation no longer feels the compact space, and the expansion rate of the 3D space is almost the same as the 4D radiation-dominated one. 

The right plot in Fig.~\ref{profile:t-hatbt} shows that a small change of $\hat{\bt}$ corresponds to a large change of $t$ in early times. 
This explains the plateau for $t\ll\trad$ in the left plot.

\item[\mbox{\boldmath $\Rmr$}-dependence] \mbox{}\\
Fig.~\ref{profile:pw3} shows the $\Rmr$-dependence of $p(t)$. 
Recall that $\Rmr$ defined in \eqref{def:cR} parameterizes the ratio of the energy density for the moduli-oscillation 
to that for the radiation at $t=\tref$. 
Since the latter energy density decreases faster than the former, 
the former eventually dominates the total energy density at late times, and $p$ will approach to $2/3$. 
The parameter~$\Rmr$ determines $\tmod$, at which the moduli-oscillation contribution starts to dominate. 
For a smaller value of $\Rmr$, it takes more time to dominate the total energy density, and thus $\tmod$ becomes larger. 
When $\Rmr>1$, on the other hand, the moduli oscillation will dominate the energy density before the universe behaves as the 4D radiation-dominated one. 

\end{description}

\subsection{Estimation of transition times} \label{transition_times}
So far, we have worked in the 6D Planck unit. 
For phenomenological discussions, however, it is more convenient to translate the physical quantities in the 4D unit. 
First, let us restore the dependence of the 6D Planck mass~$M_6$. 
\begin{align}
 t &\to \frac{t}{M_6}, \;\;\;\;\;
 \bt \to \frac{\bt}{M_6}, \;\;\;\;\;
 m_{\rm KK} = e^{-B_*} \to e^{-B_*}M_6. 
 \label{restore:M_6}
\end{align}

The 4D Planck mass~$M_4$ is defined after the extra dimensions are stabilized, and is related to the 6D Planck mass~$M_6$ as 
\begin{align}
 M_4 &\equiv \sqrt{\cV_{2*}}M_6^2 = \sqrt{4\pi}e^{B_*}M_6, 
\end{align}
where $\cV_{2*}\equiv 4\pi (e^{B_*}l_6)^2$ ($l_6\equiv M_6^{-1}$: 6D Planck length) is the volume of the compact space~$S^2$ after the moduli stabilization. 
Thus, the quantities in \eqref{restore:M_6} are expressed as
\begin{align}
 t &= \frac{\sqrt{4\pi}e^{B_*}}{M_4}t^{(6)} = 1.46\times 10^{-18}e^{B_*}t^{(6)}\,\mbox{GeV}^{-1} 
 = 8.61\times 10^{-42}e^{B_*}t^{(6)}\,\mbox{sec}, \nonumber\\
 \bt &= \frac{\sqrt{4\pi}e^{B_*}}{M_4}\bt^{(6)} = 1.46\times 10^{-18}e^{B_*}\bt^{(6)}\,\mbox{GeV}^{-1}, \nonumber\\
 m_{\rm KK} &= \frac{M_4}{\sqrt{4\pi}e^{2B_*}} = 6.87\times 10^{17}e^{-2B_*}\,\mbox{GeV}, 
\end{align}
where $t^{(6)}$ and $\bt^{(6)}$ are the values of the time and the inverse temperature measured by $M_6$. 

\begin{figure}[t]
  \begin{center}
    \includegraphics[scale=0.60]{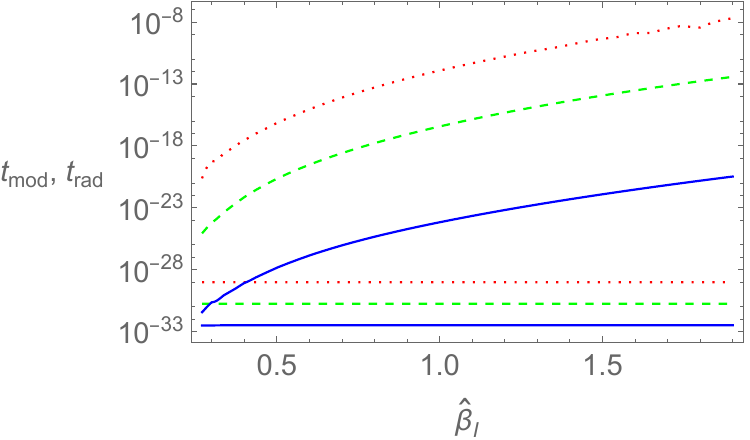} 
  \end{center}
\caption{The transition times~$\tmod$ and $\trad$ defined in \eqref{def:t_radmod} 
as functions of $\hat{\bt}_{\rm I}\equiv m_{\rm KK}\bt_{\rm I}$. 
The unit of the vertical axis is seconds. 
The solid, dashed and dotted lines correspond to the case of $\sgm_*=10$, 13 and 16, respectively. 
The upper (lower) line represents $\tmod$ ($\trad$). 
The mass parameter~$m$ is chosen as $m=0.01$. }
    \label{tmodrad-hatbtI}
\end{figure}

Fig.~\ref{tmodrad-hatbtI} shows the transition times~$\trad$ and $\tmod$ defined in \eqref{def:t_radmod} 
as functions of the initial inverse temperature normalized by the KK mass scale~$\hat{\bt}_{\rm I}\equiv m_{\rm KK}\bt_{\rm I}$. 
The solid, dashed and dotted lines correspond to the case of $\sgm_*=10$ ($m_{\rm KK}=1.2\times 10^{14}\,\mbox{GeV}$), 
13 ($6.2\times 10^{12}\,\mbox{GeV}$) and 16 ($3.1\times 10^{11}\,\mbox{GeV}$). 
From this plot, we see that $\tmod$ increases as the initial inverse temperature~$\hat{\bt}_{\rm I}$ increases. 
This can be understood by noting that the induced moduli oscillation has a larger amplitude for high initial temperature. 
Namely, for a large value of $\hat{\bt}_{\rm I}$ (i.e., low initial temperature), the pressure~$p_2^{\rm rad}$ is small 
and the induced moduli oscillation has a small amplitude, which leads to a small value of $\Rmr$. 
As shown in Fig.~\ref{profile:pw3}, this means that the moduli oscillation dominates the energy density at later time. 
In contrast, the transition from the 6D to 4D radiation-dominated eras occurs when the moduli-oscillation energy density is negligible. 
Therefore, $\trad$ is almost independent of $\hat{\bt}_{\rm I}$, as can be seen from the plot. 

\begin{figure}[t]
  \begin{center}
    \includegraphics[scale=0.60]{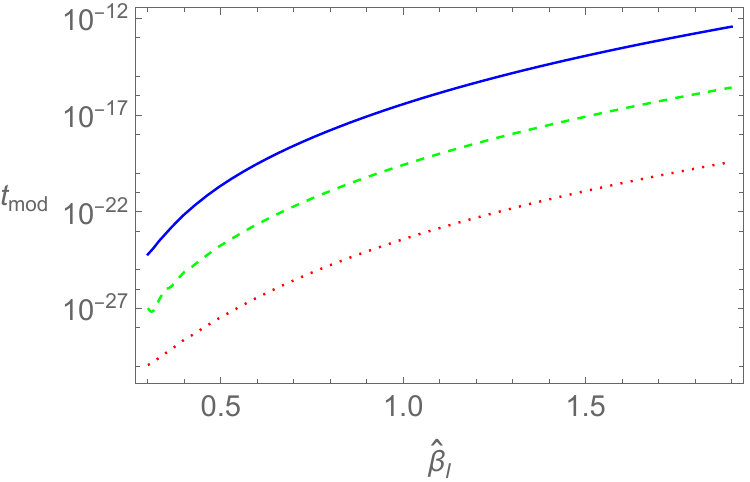} 
  \end{center}
\caption{The transition time~$\tmod$ defined in \eqref{def:t_radmod} 
as functions of $\hat{\bt}_{\rm I}\equiv m_{\rm KK}\bt_{\rm I}$ in the case of $\sgm_*=12$. 
The unit of the vertical axis is seconds. 
The solid, dashed and dotted lines correspond to the case of $m=10^{-2}$, $10^{-3}$ and $10^{-4}$, respectively. 
}
    \label{tmod-hatbtI}
\end{figure}

Next, we see the dependence of the mass parameter~$m$ in the moduli potential~\eqref{def:Vpot}. 
For a smaller value of $m$, the potential becomes shallower, and the moduli can move from the potential minimum~\eqref{static_sol} 
by the pressure~$p_2^{\rm rad}$ more easily. 
Therefore, the amplitude of the moduli oscillation becomes larger, and the value of $\Rmr$ increases. 
As a result, we have a smaller value of $\tmod$. 
This behavior can be seen in Fig.~\ref{tmod-hatbtI}, which shows the dependence of $\tmod$ on $m$. 

From \eqref{p:smallhatbt} and \eqref{fct:rho^rad-hatbt}, the transition time~$\trad$ in the $M_6$ unit is approximated as
\begin{align}
 \trad^{(6)} &\simeq \tref+\sqrt{\frac{3}{\Crad}}\brc{\cG(10)-\cG(\btref)} \simeq \sqrt{\frac{3}{\Crad}}\cG(10) 
 = \sqrt{\frac{24\pi^3}{g_{\rm dof}}}e^{3B_*}\cG(10). 
\end{align}
Since $\cG(10)\simeq 20.6$, $\trad$ in the unit of second is 
\begin{align}
 \trad &\simeq \sqrt{\frac{24\pi^3}{g_{\rm dof}}}e^{4B_*} 
 \simeq 4.83\times 10^{-39}\times\frac{e^{4B_*}}{\sqrt{g_{\rm dof}}}\;{\rm sec}. 
\end{align}

From \eqref{t-hatbt} and \eqref{def:t_radmod}, the transition time~$\tmod$ in the $M_6$ unit is approximated as
\begin{align}
 \tmod^{(6)} &\simeq \sqrt{\frac{3}{2\Crad}}\int_{\btref}^{\btmod}dx\;\frac{x^2}{v_\bt(x)\sqrt{v_\rho(x)}} \nonumber\\
 &\simeq \frac{1}{\pi^2}\sqrt{\frac{6}{\Crad}}\btmod^2 \simeq 556\sqrt{\frac{6}{\Crad}}(\Rmr)^{-2} \nonumber\\
 &\simeq 556\sqrt{6}\brkt{\frac{g_{\rm dof}e^{-6B_*}}{8\pi^3}}^{3/2}\tlCmod^{-2} 
 \simeq \frac{0.35g_{\rm dof}^{3/2}e^{-9B_*}}{\tlCmod^2}, 
\end{align}
where
\begin{align}
 \tlCmod &\equiv \Cmod e^{-3\bAref+3\cF(\btref)}. 
 \label{def:tlCmod}
\end{align}
Here, we have assumed that $\btmod\gg\btref$ and used that
\begin{align}
 \frac{x^2}{v_\bt(x)\sqrt{v_\rho(x)}} &\simeq \frac{4}{\pi^2}x, 
\end{align}
for $x\gg 1$. 

Thus, $\tmod$ in the unit of second is 
\begin{align}
 \tmod &= 8.61\times 10^{-42}e^{B_*}\tmod^{(6)} \,{\rm sec} \nonumber\\
 &\simeq \frac{3.0\times 10^{-42}g_{\rm dof}^{3/2}e^{-8B_*}}{\Cmod}e^{3\bAref-3\cF(\btref)}\,{\rm sec}. 
 \label{expr:tmod:second}
\end{align}
Since we are considering the case that $\btref\ll 1$, 
we have 
\begin{align}
 e^{3\bAref} &\simeq \brkt{1+\sqrt{\CA}\tref}^{5/3}, \;\;\;\;\;
 \btref \simeq e^{\frac{3}{5}\bAref} = \brkt{1+\sqrt{\CA}\tref}^{1/3}, 
\end{align}
from \eqref{e^A:earlytimes}. 
Plugging these and \eqref{def:Cmod} into \eqref{expr:tmod:second}, we can estimate the value of $\tmod$.

\subsection{Moduli decay}
\label{moduli_decay}
The moduli-oscillation-domination era will end by the decay of the moduli. 
After the lifetime of the moduli~$\tlf$, the moduli oscillation is converted into radiation. 
In this subsection, we will see this effect. 

From \eqref{fct:rho^rad-hatbt}, \eqref{fct:A-hatbt}, \eqref{def:cR} and \eqref{ap:calR}, the radiation energy density is written as
\begin{align}
 \rho^{\rm rad} &\simeq \Crad\frac{v_\rho(\hat{\bt})}{\hat{\bt}^6} 
 = \Crad\frac{e^{-3\cF(\hat{\bt})}}{\cR(\hat{\bt})} 
 \simeq D_{\rm ref}\frac{e^{-3\bar{A}(\hat{\bt})}}{\hat{\bt}},  
\end{align}
where
\begin{align}
 D_{\rm ref} &\equiv \frac{\Crad e^{3\bAref-3\cF(\btref)}}{0.0135}. 
\end{align}
From this, we obtain
\begin{align}
 \dot{\rho}^{\rm rad} &\simeq \brkt{-3\dot{\bar{A}}-\frac{\dot{\hat{\bt}}}{\hat{\bt}}}\rho^{\rm rad}
 \simeq -\brc{3+v_\bt(\hat{\bt})}\dot{\bar{A}}\rho^{\rm rad}, 
 \label{dotrhorad:decay}
\end{align}
where we have used \eqref{app:bt-A}. 

If we introduce the effect of the moduli decay, and \eqref{asymp:rho^mod} and \eqref{dotrhorad:decay} are modified as
\begin{align}
 \rho^{\rm mod} &\simeq \Cmod e^{-3\bar{A}-\Gmm t}, \nonumber\\
 \dot{\rho}^{\rm rad} &\simeq -\brc{3+v_\bt(\hat{\bt})}\dot{\bar{A}}\rho^{\rm rad}+\Gmm\rho^{\rm mod}, 
\end{align}
where $\Gmm\equiv 1/\tlf$ is the total decay rate of the moduli. 
Recall that 
\begin{align}
 t(\hat{\bt}) &= \tref+\int_{\btref}^{\hat{\bt}}\frac{dx}{xv_\bt(x)}\sqrt{\frac{3}{\rho^{\rm rad}(x)+\rho^{\rm mod}(x)}}, \nonumber\\
 p(\hat{\bt}) &= (t-\tc)\sqrt{\frac{\rho^{\rm rad}(\hat{\bt})+\rho^{\rm mod}(\hat{\bt})}{3}}, 
\end{align}
where $\tc$ is given by \eqref{choice:t_0}. 
Now, we numerically evaluate $p$ at each time. 
Denote the value of a quantity~$q$ at 
\begin{align}
 \hat{\bt}_i &\equiv \btref e^{\Dlt i}, \;\;\;\;\; \brkt{i=0,1,2,\cdots}
\end{align}
where $\Dlt\ll 1$ is a small positive constant, as $q_i$. 
Then, we have the following recurrence relations. 
\begin{align}
 t_{i+1} &= t_i+\dlt\tau_i, \nonumber\\
 \dlt\tau_i &\equiv \frac{\Dlt}{v_\bt(\hat{\bt}_i)}\sqrt{\frac{3}{\rho_i^{\rm rad}+\rho_i^{\rm mod}}}, \nonumber\\
 \rho_{i+1}^{\rm rad} &= \rho_i^{\rm rad}+\dlt\tau_i\sbk{-\brkt{3+v_\bt(\hat{\bt}_i)}\sqrt{\frac{\rho_i^{\rm rad}+\rho_i^{\rm mod}}{3}}\rho_i^{\rm rad}
 +\Gmm\rho_i^{\rm mod}}, \nonumber\\
 \rho_{i+1}^{\rm mod} &= \Cmod e^{-3\bar{A}_{i+1}-\Gmm t_{i+1}} \nonumber\\
 &= \rho_i^{\rm mod}\exp\brc{-\frac{\Dlt}{v_\bt(\hat{\bt}_i)}
 \brkt{3+\Gmm\sqrt{\frac{3}{\rho_i^{\rm rad}+\rho_i^{\rm mod}}}}}. 
\end{align}
At the last equality, we have used that
\begin{align}
 \bar{A}_{i+1} &= \bar{A}_i+\frac{\Dlt}{v_\bt(\hat{\bt}_i)}. 
\end{align}
Using these quantities, the effective power~$p$ at $t=t_i$ is calculated as
\begin{align}
  p_i &= (t_i-\tc)\sqrt{\frac{\rho_i^{\rm rad}+\rho_i^{\rm mod}}{3}}. 
\end{align}

\begin{figure}[t]
  \begin{center}
    \includegraphics[scale=0.60]{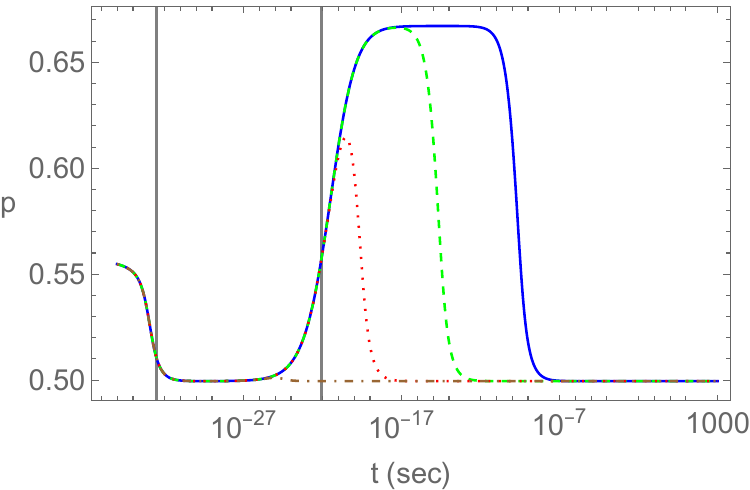} 
  \end{center}
\caption{The profile of $p(t)$ including the moduli decay effect. 
The solid, dashed, dotted and dotdashed lines correspond to the case of $\tlf=10^{-10}$, $10^{-15}$, 
$10^{-20}$ and $10^{-25}$ (sec), respectively. 
The parameters are chosen as $m=0.01$, $\sgm_*=10$ and $\bt_{\rm I}=100$ in the unit of $M_6$. 
The vertical lines denote $\trad=9.5\times 10^{-30}\,{\rm sec}$ (left) and $\tmod=2.3\times 10^{-21}\,{\rm sec}$ (right), respectively. 
}
    \label{p-t:decay}
\end{figure}

Fig.~\ref{p-t:decay} shows the effective power~$p$ as a function of $t$ in the unit of second. 
As expected, $p$ rapidly decreases to the radiation-dominated value~$1/2$ at $t=\tlf$. 

\section{Summary}
\label{conclusion}
We investigated the cosmological expansion of the 3D space in a model with two compact extra dimensions 
by solving the 6D evolution equations. 
We assumed that the whole 5D space is filled with the radiation 
and the moduli have already been stabilized at the initial time. 
In contrast to the conventional 4D effective theory analysis, the 6D evolution equations involve the pressure for the compact extra dimensions~$p_2^{\rm rad}$.  
When the temperature of the universe is higher than the compactification scale~$m_{\rm KK}$, 
the pressure~$p_2^{\rm rad}$ affects the moduli dynamics. 

In our previous work~\cite{Otsuka:2022vgf}, we found that $p_2^{\rm rad}$ pushes out the moduli 
from the potential minimum, and induces the moduli oscillation. 
If the moduli lifetime is long enough, the oscillation will eventually dominate the energy density at late times. 
In that case, the 3D space expands as $e^A\propto t^{2/3}$. 
If the temperature of the universe is higher than $m_{\rm KK}$, 
the radiation feels the whole 5D space, and the 3D space expands as $e^A\propto t^{5/9}$. 
When the temperature goes down below $m_{\rm KK}$, the radiation ceases to feel the extra dimensions, 
and the expansion rate slows down as $e^A\propto t^{1/2}$, which is the expansion law of the 4D radiation-dominated universe. 
In order to pursue these changes of the expansion rate, we define the effective power~$p$ 
in such a way that the 3D scale factor behaves as $e^A\propto t^p$ for each era. 
The nontrivial expansion of the 3D space is parameterized by two transition times~$\trad$ and $\tmod$. 
The effective power~$p$ changes from $5/9$ to $1/2$ around $t=\trad$, 
and from $1/2$ to $2/3$ around $t=\trad$.  

In our previous works~\cite{Otsuka:2022rpx,Otsuka:2022vgf}, 
we evaluated the 3D scale factor by numerical computation. 
However, it is not easy to see how the transition times~$\trad$ and $\tmod$ depend on the model parameters 
and the initial temperature in such a numerical approach. 
Besides, we cannot pursue the whole history of the universe in this method 
due to the limitation of the computational power.
In this paper, we derive analytic expressions for the 3D scale factor, the inverse temperature 
and the background moduli values by solving the 6D evolution equations under appropriate approximations, 
and provide analytic expressions for the transition times  
as functions of the model parameters~$m$, $\sgm_*$ and the initial (inverse) temperature~$\bt_{\rm I}$. 
The expressions we obtained enable us to pursue the cosmological evolution until much later times. 

The first transition time~$\trad$ is determined solely by $\sgm_*$ (or $m_{\rm KK}$), 
and is almost independent of the initial temperature. 
The second transition time~$\tmod$, on the other hand, depends on both the moduli potential scale~$m$ and the temperature. 
This is because $\tmod$ is determined by the oscillation amplitude induced by $p_2^{\rm rad}$. 
The amplitude becomes larger for a shallower potential (i.e., a smaller value of $m$ or a larger value of $\sgm_*$) 
or for higher initial temperature (i.e., a smaller value of $\bt_{\rm I}$), 
and then the moduli oscillation dominates the energy density earlier (a smaller value of $\tmod$). 

As shown in Ref.~\cite{Otsuka:2022rpx}, 
the modulus~$B$ continues to increase for $\sgm_*\gtrsim 16$, and the observed 4D universe cannot be obtained. 
Therefore, there is an upper bound for the stabilized value of the $S^2$ radius in our model. 
It is intriguing to study whether this is common to other models with extra dimensions or not. 

For more realistic discussions, we need to extend our setup by including the inflaton sector. 
Our initial conditions in \eqref{initial_conds} with the radiation-domination assumption 
should be realized by the reheating process after the inflation. 
In such setups, the effective power~$p$ will enter the expression of the e-folding number for the 3D scale factor. 

We will discuss these issues in separate papers.

\appendix

\section{Thermodynamic quantities}
\label{TDquantities}
The dispersion relation of a 6D relativistic or massless particle is 
\begin{align}
k^Mk_M &= -k_0^2+e^{-2A}\vec{k}^2+e^{-2B}k_\theta^2+\frac{1}{e^{2B}\sin^2\theta}k_\phi^2 = 0. 
\end{align}
Thus the energy of the particle with the 3D momentum~$\vec{k}=(k_1,k_2,k_3)$ and the angular momentum~$l$ on $S^2$ is given by
\begin{align}
 {\cal E}_{k,l} &= k_0 = \sqrt{e^{-2A}k^2+e^{-2B}l(l+1)}, 
\end{align}
where $k\equiv \sqrt{\vec{k}^2}$. 
Since each one-particle state is specified by $\vec{k}$, $l$ and the `magnetic quantum number'~$m=-l,\cdots,l$, 
we have $(2l+1)$ degenerate energy eigenstates for each $\vec{k}$ and $l$. 
Hence, the grand potential is expressed as 
\begin{align}
 J(\beta,\mu,{\cal V}_3,{\cal V}_2) &= \pm\sum_{l=0}^\infty\frac{g_{\rm dof}(2l+1)}{2\pi^2\beta}
 \int_0^\infty dk\;k^2\ln\brkt{1\mp e^{-\beta({\cal E}_{k,l}-\mu)}} \nonumber\\
 &= \mp\frac{g_{\rm dof}{\cal V}_3}{\pi^2\beta^4}{\rm Li}_4(\pm e^{\beta\mu})
 \pm\sum_{l=1}^\infty\frac{g_{\rm dof}(2l+1){\cal V}_3}{2\pi^2\beta^4}
 \int_0^\infty dq\;q^2\ln\brkt{1\mp e^{-\sqrt{q^2+c_l^2}+\beta\mu}}, 
 \label{def:J}
\end{align}
where $g_{\rm dof}$ denotes the degrees of freedom for the 6D relativistic particles, $\beta$ is the inverse temperature, 
$\mu$ is the chemical potential, and ${\cal V}_3\equiv e^{3A}$ and ${\cal V}_2\equiv4\pi e^{2B}$ are the comoving volume for the 3D space 
and the physical volume of $S^2$, respectively. 
The upper (lower) signs correspond to the case of bosons (fermions). 
At the second equality, we have rescaled the integration variable and the KK masses as 
\begin{align}
 q &\equiv e^{-A}\bt k, \;\;\;\;\;
 c_l \equiv \beta\sqrt{\frac{4\pi l(l+1)}{{\cal V}_2}} = e^{-B}\bt\sqrt{l(l+1)}. 
\end{align}
The function~${\rm Li}_4(z)$ in the second line of (\ref{def:J}) is the polylogarithmic function. 
In the following, we consider a situation in which $e^{-c_l+\beta\mu}\ll 1$ for $l\geq 1$. 
Then the grand potential can be approximated as
\begin{align}
 J(\beta,\mu,{\cal V}_3,{\cal V}_2) &\simeq -\frac{g_{\rm dof}{\cal V}_3}{2\pi^2\beta^4}
 \brc{\pm 2{\rm Li}(\pm e^{\beta\mu})+e^{\beta\mu}Q_1\brkt{\beta\sqrt{\frac{4\pi}{{\cal V}_2}}}}, 
\end{align}
where
\begin{align}
 Q_1(x) &\equiv \sum_{l=1}^\infty x^2l(l+1)(2l+1)K_2\brkt{x\sqrt{l(l+1)}}. 
 \label{def:Q_1}
\end{align}
Here, $K_2(z)$ is the modified Bessel function of the second kind. 

From (\ref{def:J}), various thermodynamic quantities are calculated as follows. 
\begin{description}
\item[Radiation energy density] 
\begin{align}
 \rho^{\rm rad} &= \frac{1}{{\cal V}_3{\cal V}_2}\brkt{\partial_\beta-\frac{\mu}{\beta}\partial_\mu}(\beta J) \nonumber\\
 &\simeq \frac{g_{\rm dof}}{2\pi^2\beta^4{\cal V}_2}\brc{\pm 6{\rm Li}_4(\pm e^{\beta\mu})+e^{\beta\mu}\brkt{3Q_1+Q_2}}, 
 \label{expr:rho^rad}
\end{align}
where 
\begin{align}
 Q_2(x) &\equiv -xQ_1'(x) = \sum_{l=1}^\infty x^3l^{3/2}(l+1)^{3/2}(2l+1)K_1\brkt{x\sqrt{l(l+1)}}. 
 \label{def:Q_2}
\end{align}

\item[3D pressure]
\begin{align}
 p^{\rm rad}_3 &= -\frac{1}{{\cal V}_2}\frac{\partial J}{\partial {\cal V}_3}
 \simeq \frac{g_{\rm dof}}{2\pi^2\beta^4{\cal V}_2}\brc{\pm 2{\rm Li}_4(\pm e^{\beta\mu})+e^{\beta\mu}Q_1}. 
 \label{expr:p^rad_3}
\end{align}

\item[2D pressure]
\begin{align}
 p^{\rm rad}_2 &= -\frac{1}{{\cal V}_3}\frac{\partial J}{\partial {\cal V}_2}
 \simeq \frac{g_{\rm dof}e^{\beta\mu}}{4\pi^2\beta^4{\cal V}_2}Q_2. 
 \label{expr:p^rad_2}
\end{align}

\end{description}
The arguments of the functions~$Q_1$ and $Q_2$ are understood as 
$\beta\sqrt{4\pi/{\cal V}_2}=e^{-B}\beta$. 

We should note that
\begin{align}
 \rho^{\rm rad} &= 3p^{\rm rad}_3+2p^{\rm rad}_2. \label{rel:rhop}
\end{align}

In this paper, we assume that $\beta\mu\ll 1$ and neglect the chemical potential~$\mu$. 
Note that 
\begin{align}
 {\rm Li}_4(1) &= \zeta(4) = \frac{\pi^4}{90}, \;\;\;\;\;
 {\rm Li}_4(-1) = -\frac{7}{8}\zeta(4). 
\end{align}
If we have the same degrees of freedom for bosons and fermions, the total energy density and pressures are expressed as
\begin{align}
 \rho^{\rm rad} &\simeq \frac{g_{\rm dof}}{2\pi^2\bt^4\cV_2}\brc{\frac{\pi^4}{16}+3Q_1(e^{-B}\bt)+Q_2(e^{-B}\bt)}, \nonumber\\
 p_3^{\rm rad} &\simeq \frac{g_{\rm dof}}{2\pi^2\bt^4\cV_2}\brc{\frac{\pi^4}{48}+Q_1(e^{-B}\bt)}, \nonumber\\
 p_2^{\rm rad} &\simeq \frac{g_{\rm dof}}{4\pi^2\cV_2}Q_2(e^{-B}\bt), 
 \label{expr:rho-p}
\end{align}
where $g_{\rm dof}$ is the total degrees of freedom for the radiation.

\section{Conservation law}
\label{conserv_law}
Including the radiation contribution, the energy-momentum conservation law is 
\begin{align}
 \nabla_M T^M_{\;\;N} &\equiv \partial_M T^M_{\;\;N}+\Gamma^M_{\;\;ML}T^L_{\;\;N}-\Gamma^L_{\;\;MN}T^M_{\;\;L} = 0, 
 \label{T:conserve}
\end{align}
where 
\begin{align}
 T^t_{\;\;t} &= \frac{1}{2}\dot{\sigma}^2+\frac{e^\sigma}{8b^4}+V(\sigma)+\rho^{\rm rad} \equiv \rho^{\rm tot}, \nonumber\\
 T^i_{\;\;j} &= \delta^i_{\;\;j}\brc{-\frac{1}{2}\dot{\sigma}^2+\frac{e^\sigma}{8b^4}+V(\sigma)-p^{\rm rad}_3} 
 \equiv -\delta^i_{\;\;j}p^{\rm tot}_3, \nonumber\\
 T^4_{\;\;4} &= T^5_{\;\;5} = -\frac{1}{2}\dot{\sigma}^2-\frac{e^\sigma}{8b^4}+V(\sigma)-p^{\rm rad}_2
 \equiv -p^{\rm tot}_2. 
 \label{comp:T}
\end{align}
From (\ref{T:conserve}) with $N=t$, we have 
\begin{align}
 \dot{\rho}^{\rm tot}+3\dot{A}\brkt{\rho^{\rm tot}+p^{\rm tot}_3}+2\dot{B}\brkt{\rho^{\rm tot}+p^{\rm tot}_2} &= 0, 
 \label{rho:conserve}
\end{align}
where the dot denotes the time derivative. 
The other components hold trivially. 
By using the dilaton field equation, the conservation law~(\ref{rho:conserve}) is reduced to
\begin{align}
 \dot{\rho}^{\rm rad}+\brkt{3\dot{A}+2\dot{B}}\rho^{\rm rad}
 +3\dot{A}p^{\rm rad}_3+2\dot{B}p^{\rm rad}_2 &= 0. 
 \label{radiation:conserve}
\end{align}
Plugging \eqref{expr:rho-p} into this, we obtain 
\begin{align}
 \frac{\dot{\beta}}{\beta}\brkt{\frac{\pi^4}{4}+12Q_1+5Q_2+Q_3} 
 &= 3\dot{A}\brkt{\frac{\pi^4}{12}+4Q_1+Q_2}
 +\dot{B}\brkt{2Q_2+Q_3}, 
 \label{eq_for_beta}
\end{align}
where 
\begin{align}
 Q_3(x) &\equiv 2Q_2(x)-xQ_2'(x) \nonumber\\
 &= \sum_{l=1}^\infty x^4l^2(l+1)^2(2l+1)K_0\brkt{x\sqrt{l(l+1)}}. 
 \label{def:Q_3}
\end{align}
The arguments of $Q_{1,2,3}$ are understood as $e^{-B}\beta$. 

\begin{figure}[t]
  \begin{center}
    \includegraphics[scale=0.40]{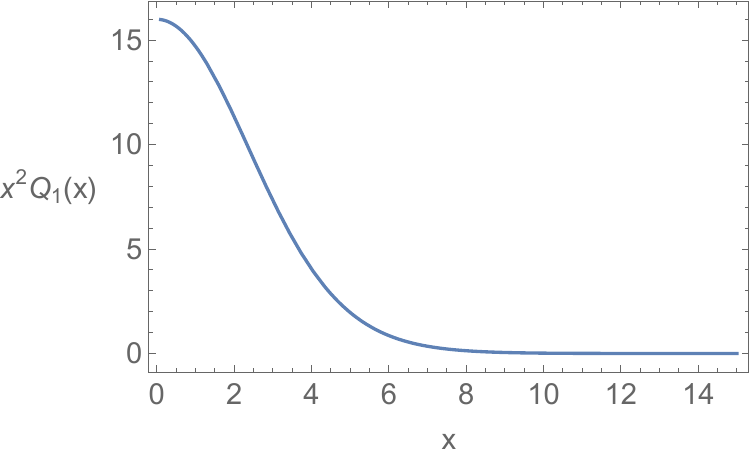} \;\;
    \includegraphics[scale=0.40]{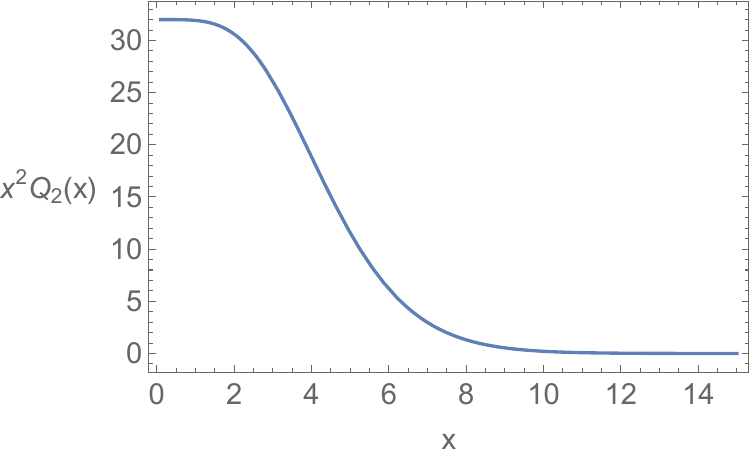} \;\;
    \includegraphics[scale=0.40]{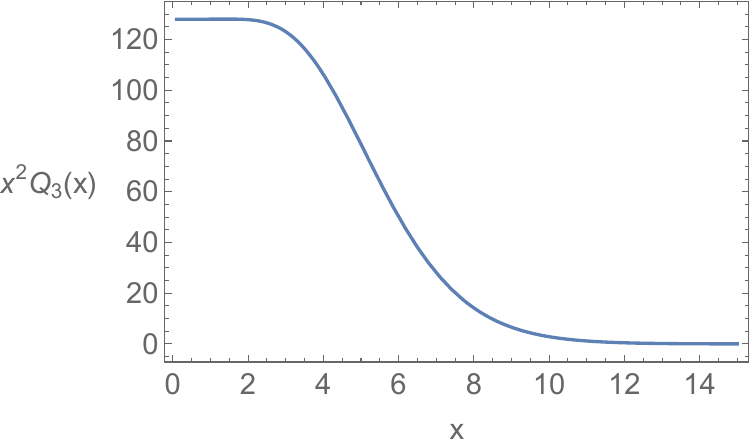}
  \end{center}
\caption{The profile of $x^2Q_1(x)$, $x^2Q_2(x)$ and $x^2Q_3(x)$. 
}
    \label{Q2}
\end{figure}

Fig.~\ref{Q2} shows the profiles of $Q_{1,2,3}(x)$. 
For $x\ll 1$, they are approximated as
\begin{align}
 Q_1(x) &\simeq \frac{16}{x^2}, \;\;\;\;\;
 Q_2(x) \simeq \frac{32}{x^2}, \;\;\;\;\;
 Q_3(x) \simeq \frac{128}{x^2}. 
 \label{Qs:smallx}
\end{align}

\section{Induced oscillation solution}
\label{forced_osc}
Here, we provide a solution of the following equation for a real function~$\vph(t)$. 
\begin{align}
 \ddot{\vph} &= -\lmd\vph+\frac{\alp}{\brkt{1+\sqrt{C}t}^q}, 
\end{align}
where $\lmd>0$, $C>0$, $\alp$ and $q$ are real constants. 
The solution is expressed as
\begin{align}
 \vph(t) &= c_1\cos\brkt{\sqrt{\lmd}t}+c_2\sin\brkt{\sqrt{\lmd}t}-\frac{\alp}{\lmd}\brkt{\frac{\lmd}{C}}^{\frac{q}{2}}
 \cS\brkt{\sqrt{\frac{\lmd}{C}}\brkt{1+\sqrt{C}t}}, 
\end{align}
where $c_1$ and $c_2$ are integration constants, and 
\begin{align}
 \cS(x) &\equiv \Im\brc{e^{(q-1)\frac{\pi}{2}i}\cE_q(ix)}, \nonumber\\
 \cE_q(x) &\equiv e^z\frac{{\rm E}_q(z)}{z^{q-1}} = e^z\Gm(1-q,z). 
\end{align}
The function~${\rm E}_q(z)\equiv \int_1^\infty dw\;e^{-zw}/w^q$ is the (generalized) exponential integral, 
and $\Gm(s,z)$ is the upper incomplete gamma function. 
The derivatives of the real function~$\cS(x)$ are 
\begin{align}
 \cS'(x) &= \Re\brc{e^{(q-1)\frac{\pi}{2}i}\cE'_q(ix)} = \Re\brc{e^{(q-1)\frac{\pi}{2}i}\cE_q(ix)}, \nonumber\\
 \cS''(x) &= -\Im\brc{e^{(q-1)\frac{\pi}{2}i}\cE'_q(ix)} = -\frac{1}{x^q}-\cS(x). 
\end{align}
We have used that
\begin{align}
 \cE_q'(z) &= \frac{-1+e^z z{\rm E}_q(z)}{z^q} = -\frac{1}{z^q}+e^z\Gm(1-q,z) = -\frac{1}{z^q}+\cE_q(z). 
\end{align}

When the initial condition is that $\vph(0)=\dot{\vph}(0)=0$, the solution becomes 
\begin{align}
 \vph(t) &= -\frac{\alp}{\lmd}\brkt{\frac{\lmd}{C}}^{\frac{q}{2}}\Biggl\{
 \cS\brkt{\sqrt{\frac{\lmd}{C}}\brkt{1+\sqrt{C}t}} \nonumber\\
 &\hspace{25mm}
 -\cS\brkt{\sqrt{\frac{\lmd}{C}}}\cos\brkt{\sqrt{\lmd}t}
 -\cS'\brkt{\sqrt{\frac{\lmd}{C}}}\sin\brkt{\sqrt{\lmd}t}\Biggr\} \nonumber\\
 &= -\frac{\alp}{\lmd}\brkt{\frac{\lmd}{C}}^{\frac{q}{2}}\Im\brc{e^{(q-1)\frac{\pi}{2}i}\cU_q(t;\lmd,C)}, 
 \label{sol:vph}
\end{align}
where
\begin{align}
 \cU_q(t;\lmd,C) &\equiv \cE_q\brkt{i\sqrt{\frac{\lmd}{C}}\brkt{1+\sqrt{C}t}}
 -\cE_q\brkt{i\sqrt{\frac{\lmd}{C}}}e^{i\sqrt{\lmd}t} \nonumber\\
 &= e^{i\sqrt{\frac{\lmd}{C}}\brkt{1+\sqrt{C}t}}
 \brc{\Gm\brkt{1-q,i\sqrt{\frac{\lmd}{C}}\brkt{1+\sqrt{C}t}}-\Gm\brkt{1-q,i\sqrt{\frac{\lmd}{C}}}}. 
 \label{def:cU}
\end{align}
The derivative of \eqref{sol:vph} is expressed as
\begin{align}
 \dot{\vph} &= -\alp\brkt{\frac{\lmd}{C}}^{\frac{q}{2}}\Re\brc{e^{(q-1)\frac{\pi}{2}i}\cU_q(t;\lmd,C)}. 
 \label{dotvph}
\end{align}

\printbibliography

\end{document}